# Polarization-dependent photocurrent in a quadrilateral-shaped bulk crystalline tellurium chip with near-infrared light excitation

Hiro Munekata[1]*, Gakuto Kusuno[1], Kohei Miyazaki[1], and Takuya Satoh[1,2]*

[1] *Department of Physics, Institute of Science Tokyo, Tokyo 152-8551, Japan*

[2] *Quantum Research Center for Chirality, Institute for Molecular Science, Okazaki 444-8585, Japan*

* E-mail: munekata.h.aa@m.titech.ac.jp; satoh@phys.sci.isct.ac.jp

**ABSTRACT**

We report the detection of zero-bias photocurrents induced by interband excitation using oblique near-infrared light ($\lambda$ = 0.780 μm) along the directions parallel and perpendicular to the helical axis in a millimeter-sized Te crystal. The photocurrent parallel to the helical axis exhibits a circular-polarization-dependent component, namely circular photogalvanic effect. We create a framework for estimating the gyrotropic photoconductivity tensor $\beta$, the ratio between the circular-polarization-dependent photocurrent density and intensity of light, through which extrinsic $\beta$ values ranging from a few to a few tens of nA/W are obtained. Searching the $\beta$ values through existing literatures reveals that these values vary significantly with sample conditions, excitation wavelengths, and forms of crystallite samples. A particular interesting point found in our study is that the $\beta$ values for the interband excitation may be greater than those for intraband excitation as far as bulk Te is concerned.

## 1. Introduction

### 1.1 Background

Let us imagine a spiral wire, along which tellurium (Te) atoms are placed regularly in such a way that the fourth Te atom lies directly on the first Te atom, namely, a long, trigonal wire made of Te atoms [Fig. 1(a)].[1] The number of such wires condenses into a hexagonal array [Fig. 1(b)], with the axis of the spiral wire, *the helical axis*, parallel to the direction of the pillar of a hexagonal prism, the $z$ axis [Fig. 1(c)]. Although the Te crystals take the form of hexagonal prisms, they have a threefold symmetry, as shown in Fig. 1(b).

Experimental studies on the electronic and optical properties and device applications of Te crystals and films were conducted primarily in the 1960s to 1970s, with the central motivation being the elucidation of the energy band structure, understanding its exotic physical properties, and its applications in solid-state devices. Representative experimental studies include, but are not limited to, infrared absorption spectroscopy,[2–5] cyclotron resonance,[6] and carrier transport.[7-9] These studies have established that a band gap is around 0.33 eV with rather weak temperature dependence. Te crystals exhibit anisotropic carrier transport with p-type conduction at low temperatures, whereas the state of intrinsic conduction takes over at 250 K or higher.[8,9] Thin-film transistors have been tested,[10,11] and the fundamental concept of optical memory has been developed through the studies using Te-based films.[12–14]

The work with the tight binding scheme along a spiral wire have led to theoretical consideration of energy bands for this type of crystal.[15] Group theoretical methods were introduced to account for energy spectra in connection with a few different symmetry points in the Brillouin zone.[16] These two studies were followed by calculations of energy bands with various theoretical treatments in 1966 and 1967, through which the presence of an energy gap around the $H$ point of the Brillouin zone was clarified. The remaining problem of valence band structures, either single maximum or double maxima, was finally clarified by very careful calculations using the group theory and $\mathbf{k}\cdot\mathbf{p}$ perturbation method[17–19] involving severe comparison with experimental results.[4–6] Based on these theoretical studies, the double-band maximum accompanied by *non-degenerate* spin states, in other words, *non-degenerate* states of the total angular momentum, has finally been established for the top part of the valence band. Critical reviews of previous theoretical studies are presented in Refs. 17.

A new photogalvanic effect (PGE) induced by optical excitation with circularly polarized light, the circular PGE (CPGE), in a chiral crystal was suggested in the late 70's,[20] presumably after careful study of the theoretical studies mentioned in the previous paragraph. The experiments were performed to determine photo-induced electromagnetic force (emf) across crystalline Te rods using circularly polarized, pulsed infrared light excitation.[21,22] The emf of 0.2−1.0 mV was observed along the $z$ axis with 100-ns pulse excitation and peak power of 3 kW at 10.6 μm wavelength of a Q-switch $CO_2$ laser. They considered a microscopic mechanism in terms of virtual intraband excitation as described below.

Figure 2 schematically illustrates the band structure of Te and its excitation mechanisms. Process A represents the direct absorption of spin +1 photons ($A_1$) and indirect emission of spin +1 photons ($A_2$) between the conduction band (CB) and uppermost valence band ($VB_1$), whereas process A' involves direct absorption ($A'_1$) and indirect emission ($A'_2$) of spin +1 photons between the CB and lower second valence band ($VB_2$). The difference in photon energy, $h\nu = h\nu(A_1) - h\nu(A_2) = h\nu(A'_1) - h\nu(A'_2)$, corresponds to the energy of the virtual intraband excitation. These two processes break the balance of the carrier population between the two opposite spins in $K$-space, thus yielding a CPGE. The imbalance of electron spins (the light-brown zone in Fig. 2) results in a CPGE at room temperature, whereas an imbalance of hole spins (the light-blue zone in Fig. 2) dominates at low temperatures.[22]

Interest in searching for a Weyl semimetal, a system incorporating three-dimensional Dirac cones without spin degeneracy around the Fermi level,[24] has led theoretical researchers to shed light again on Te and Se crystals.[25,26] Furthermore, experimental studies have been invoked in view of modern spin-related phenomena, as represented by current-induced magnetization in a bulk Te crystal,[27,28] electrical magnetochiral anisotropy.[29] Studies of zero-bias photocurrents in view of PGE have been theoretically[30] and experimentally conducted in Te flakes[31-33] and bulk crystals.[34]

## 1.2 Objectives

Through studying the interaction between light and solids, more specifically, the interaction between circularly polarized light and spin/phonon subsystems,[35–37] we developed a strong interest in chiral systems because they would couple relatively strongly with the circular polarization (CP) of light. Indeed, we have reported the experimental evidence of such coupling, truly chiral phonons in bulk Te and HgS using circularly polarized Raman spectroscopy.[38,39] We infer that these types of crystals would also be suitable for studying the photogeneration of a spin-polarized current, as well as CP photodetectors[40] and emitters.[41]

The next step involves setting up a stage for experimentally studying the PGE. We started the study in 2022 with bulk Te crystals because they are available on the market, and a rich experimental and theoretical knowledge has been accumulated. Based on the perspective of sample preparation, we have thought that a bulk form, after it's crystal quality and orientation are clearly characterized, allows us to apply multiple electrode pairs that enable photocurrent measurements along various directions. By now, the occurrence of CPGE via the intraband excitations (Fig. 2) using polarized infrared/terahertz radiation ($\lambda$ = 9.6 − 270 μm) is established by the observation and analysis of polarization-dependent photocurrents in bulk Te crystals[34] and Te flakes.[31] Furthermore, the gyrotropic tensor $\beta$,[42] the ratio between circular-polarization-dependent photocurrent density and intensity of light, has been calculated in the intrinsic case: it is around 1.5 μA/W at room temperature.[30] A question is the extent to which interband excitation, illustrated by the blue arrow B in Fig. 2, influences the PGE, given that carrier relaxation dynamics are much more pronounced under the interband excitation. Partial answers emerged in 2023 from experiments with Te flakes.[32] In the present study, we report circular-polarization-dependent photocurrents induced by oblique near-infrared light ($\lambda$ = 0.780 μm, $h\nu$ = 1.59 eV) along the direction parallel to the z axis in millimeter-sized Te crystals.

The remainder of this article is organized as follows. In the next section, the crystallographic evaluation of cleaved Te mother pieces using X-ray diffraction is described. Furthermore, the estimation of the carrier concentrations is presented. In the third section, the device fabrication steps with pairs of electrodes parallel and perpendicular to the helical axis are described. Measures to prevent unforeseen deterioration in the photocurrents are also discussed. Section 4 presents the evaluation of the sample surfaces using Raman scattering, including identification of the chirality of the mother pieces. Section 5 describes the experimental setup for photocurrent measurements and introduces the model used to convert the measured photocurrent into photocurrent density in this study.

The sixth section is dedicated to the experimental results of anisotropic photocurrents, followed by an analysis of the experimental data using phenomenological photocurrent equations. The aim is to estimate the extrinsic $\beta$ value under interband excitation and to examine how extrinsic factors, such as crystal quality and device fabrication, influence this parameter. In the absence of clear precedents in the literature, we developed a dedicated framework for this purpose, which is described in sub-section 6.3. Hence, the extracted extrinsic $\beta$ values vary with sample conditions, ranging from a few nA/W to several tens of nA/W.

## 2. Materials

A Te lump (purity 4N) was first cleaved into 5–15-mm large, 0.5–2-mm thick mother pieces. The pieces were then characterized using a multi-purpose $\theta$-2$\theta$ X-ray diffractometer (D8 DISCOVER, Bruker; X-ray source Cu-K$\alpha$ 1.541 Å), and in-plane and out-of-plane crystal axes were identified. The area of X-ray irradiation on the sample is estimated to be approximately 1.9 mm × 0.4 mm at the Bragg's angle of $\theta_B \sim 22°$.[43] It is worth noting that X-ray diffraction (XRD) detects a Te crystal as a crystal with *six-fold symmetry*, where spacings and directions of periodic reflection planes; namely, {100} and {010}, are equivalent. Reflection planes and their directions are specified on the basis of Miller indices, as shown in Fig. 1(c): namely, **a** ∥ [100], **b** ∥ [010], and **c** ∥ [001], whereas the direction normal to the (010) plane is **y** ∥ [120]. Superimposing the orthogonal crystal coordinate system, we see [100] ∥ **x**, [120] ∥ **y**, and [001] ∥ **z**, respectively.

Figures 3(a) and (b) show two pole figures obtained at the Bragg's angles of $2\theta_B$ = 22.8–23.4 and 27.4–28.0°, respectively,[43] for one of the mother pieces [Fig. 3(c)]. The former is the reflection from the {010} plane, whereas the latter is the reflection from the {011} plane. Shown in Fig. 3(c) is a top view of the mother Te piece, on which an X-ray beam is impinged from the three o'clock direction ($\varphi = 0°$) with an incident angle $\omega$ ($\equiv \theta_B$). Symbols $\varphi$ and $\psi$ in the pole figures represent the in-plane rotation and tilt angles of the sample stage, respectively, with respect to the incident X-ray plane. The measurement configurations for $\varphi = 0°$ and 90° are shown schematically in Figs. 3(d) and (e), respectively.

The elongated bright spot centered at ($\varphi$, $\psi$) = (0, 0), spot *a* in Fig. 3(a), indicates that the (010) plane lies parallel to the sample surface, which is consistent with previous studies on cleaved planes.[1,44] The elongation along the radial direction indicates that the reflection from the (010) plane involves a dispersion of approximately 10° along the direction orthogonal to the plane of X-ray incidence ($\varphi$ = 0°). Spots *b* and *c* in Fig. 3(a) are the reflection obtained from the neighboring (010) plane, namely, ($\bar{1}$10) and (100), respectively.

Rotating and tilting the sample at the Bragg's angle of $2\theta_{XRD}$ = 27.7° give rise to the observation of a series of refection from the {011} planes; e.g., representative spots, *d*, *e*, and *f* at ($\varphi$, $\psi$) = (270°, 65°), (37°, 65°), and (127°, 65°) in Fig.3(b) correspond to the reflection from (011), ($\bar{1}$11), and (101) planes, respectively. The observed pattern indicates that the [001] direction is along $\varphi$ = 0° which is parallel to the X-ray incidence plane. The elongation in these spots indicates a dispersion of approximately 10° along the direction orthogonal to the plane of X-ray incidence ($\varphi = 90°$); for instance, the 10° dispersion of the [001] direction in the case of spot *d*.

The XRD data from different points on the mother Te chip qualitatively exhibited the same patterns as those shown in Figs. 3(a) and (b). To summarize these two pole figures, the direction normal to a sample surface is [120] ∥ **y**, whereas the direction parallel to a long side is [001] ∥ **z**. Naturally, the direction parallel to a short side is [100] ∥ **x**. Our materials consist of highly oriented crystalline textures, which, although not ideal single crystals, still allow the reliable investigation of photogalvanic effects. Dispersion of around 10° indicates disorder of the crystalline planes in terms of macroscopic orientation and microscopic spatial distance between the planes. The in-plane and out-of-plane distributions of these disorders have not yet been clarified.

As stated in the introduction, Te crystals exhibit p-type conduction at low temperatures,[8,9] whereas intrinsic conduction takes over above 250 K, which results in a sign reversal of the Hall coefficient from positive to negative. This fact indicates that the carrier concentrations estimated from the Hall coefficients at relatively high temperatures are just nominal values. In any case, we suppose that electrons and holes coexist in our samples that are not intentionally doped: for instance, the intrinsic electron/hole concentration at room temperature is estimated to be around $1.5 \times 10^{15}$ $cm^{-3}$.[45] It is thus reasonable to assume that the Fermi level is around the middle of the bandgap, as shown in Fig. 2.

## 3. Device Fabrication

Device fabrication of exotic crystalline materials is often difficult. We developed a technique that is convenient and free from chemical processing. Mother pieces that clearly exhibited an in-plane *z* axis were selected for device fabrication. Selected pieces were then cut into quadrilateral-shaped, millimeter-sized Te chips with great care to take on near-rectangular shapes, with their sides either parallel or perpendicular to the *z* axis. This process was carried out under a stereomicroscope using jewelry processing tools such as a microstring saw and a microfile. A successful example is shown in Fig. 4(a). During the CPGE experiments with the *as-cleaved* devices, we observed deterioration in the polarization-dependent photocurrent. This prompted us to examine the suppression of the deterioration concurrently with CPGE. It has been inferred that photoinduced oxidation and subsequent amorphization of the crystalline Te surface occurred.[46] Hence, devices with mirror-polished and $SiO_2$-coated surfaces were prepared in addition to those with *as-cleaved* surfaces. The process of surface polish and $SiO_2$-coating, which is relatively simple and free from conventional chemical lithography, is described in the next paragraph.

Figures 4(b) and (c) show schematic illustrations of the device from the top and cross-sectional views, respectively. First, the chip was mounted on a cover glass using an epoxy-based adhesive, with the as-cleaved {010} plane facing upward. The area surrounded by the red lines in Figs. 4(b) and (c) represent the epoxy area. Second, the sample surface was mirror-polished using a mixture of alumina powder and nonaqueous solvent, and the resultant sample thickness was unified at approximately 300 μm. Alumina powders of 3, 1, and 0.3 μm in diameter were used. No post-polishing chemical etching was performed. This was followed by electrode formation: metal fingers were applied on each side of a chip with silver paste or a mixture of gold-silver paste, and aluminum (Al) wires of 100-μm diameter were bonded to the ends of the fingers. Two- and four-finger devices were fabricated. It has been reported that, around room temperature, no noticeable solid reaction occurs between Ag (Au) and Te.[47)] We also tested the stability of contact resistance by intentionally illuminating Ag/Te boundaries of devices for 30 min. or longer, which resulted in no change in the resistance. Finally, some of the as-cleaved devices were coated with a 300-nm-thick amorphous $SiO_2$ layer by RF sputtering.[48)] The regions surrounded by the green lines correspond to the $SiO_2$ layer. Photographs of the actual two- and four-terminal devices are presented in the insets of Figs. 5(a) and 6(a), respectively. For the devices with *as-cleaved* surfaces, the second and final steps were omitted. Two representative experimental results obtained from the two-finger device (Sample A) and four-finger device (Sample B) are presented in Sections 6.1 and 6.2, respectively. Current-voltage data of these devices are exemplified in Fig. 1S in the Supplementary Information.

## 4. Raman Scattering

Raman scattering characterization was carried out to study the possible damage caused by RF sputtering of $SiO_2$-coated devices. Furthermore, light-induced oxidation[46)] were also tested experimentally. Raman spectra were measured in a standard backscattering configuration at room temperature in an air atmosphere using a semiconductor laser with a wavelength $\lambda$ of 785 nm and a power of 7 mW (density 50 kW/cm$^2$). Figure 5(a) shows the Raman spectra obtained from the *$SiO_2$-coated* sample A. Depending on the Raman excitation selection rule, the two lines at 124 and 144 $cm^{-1}$ are clearly identified as $A_1$ and $E$ phonon modes.[46,49] The full width at half maximum of these lines are 8.5 ± 0.1 $cm^{-1}$ (HH) and 10.2 ± 0.9 $cm^{-1}$ (VH) for the $A_1$ mode, whereas 9.2 ± 1.2 $cm^{-1}$ (HH) and 7.6 ± 0.6 $cm^{-1}$ (VH) for the $E$ mode, which are comparable to those observed in carefully treated Te crystals.[49] This indicates that the device surface is not severely damaged by the process of $SiO_2$ coating. Figure 5(b) shows Raman data in the relatively high wavenumber region (400–800 $cm^{-1}$). This region is important because early researchers have reported the production of $TeO_x$ layers in the regions represented by yellow zones in the figure, during the illumination of a CW-laser beam ($\lambda$ = 441.6, 488.0, and 514.5 nm, 0.1–100 kW/$cm^{-2}$) in an air atmosphere.[46] Fortunately, the occurrence of the oxides was not observed within the limit of illumination at $\lambda$ = 785 nm and exposure time up to 92 min. It appears that the $SiO_2$ layer prevented the light-induced oxidation and subsequent amorphization of the Te underlayer region.

It is important to correlate the chirality of the samples with their physical properties. We attempted to identify chirality using circular-polarized Raman scattering,[38] in which a light beam was aligned parallel to [001] axis to detect chirality-dependent lines. We applied this method for the Te mother pieces and found that they showed the circular-polarization-dependent splitting in the $\Gamma_3$ doublet, from which pieces were identified as right-handed Te ($P3_121$).

## 5. Photocurrent Measurement

A sample was mounted on a vertical sample rod in an aluminum sample box in such a way that [120] ∥ **y** direction was parallel to the sample normal and [001] ∥ **z** direction pointed horizontally. Naturally, [100] ∥ **x** direction pointed vertically. The sample was rotated around the [100] axis at a tilt angle $\theta$ relative to the [120] direction. A picture of sample B mounted in the box is shown in Fig. 6(a), along with a schematic illustration of the relationship between the direction of the Miller indices and the orthogonal coordinate system ($x$, $y$, $z$). Hereafter, the latter coordinate system is used primarily to describe the experimental results and discussions in subsequent sections.

Excitation was carried out with a CW semiconductor laser beam at $\lambda$ = 780 nm and a power density in the range of 170–4400 W/$cm^2$, which was controlled by the output power of the laser, neutral-density filters, and spot diameters of either approximately 40 or 100 μm. Details of excitation are specified in the experimental data presented in the next section. The linearity of the current–voltage curves and response under light illumination at different spots with various excitation conditions were tested for a few hours with normal incidence ($\theta = 0°$) prior to the experiments on PGE.

A schematic diagram of the experimental setup is shown in Fig. 6(b). The penetration depth of light was estimated to be 20−50 nm at room temperature referring to the absorption spectra measured by earlier researchers.[3] A zero-bias photocurrent was measured at room temperature in an air atmosphere with different tilt angles $\theta$. A lock-in technique using a beam chopper operating at 163 Hz was employed. A quarter wave plate (QWP) was rotated continuously during measurements, and a photocurrent as a function of rotation angle ($\varphi_Q$) of QWP was obtained. The linear polarizations (LPs) of a laser beam parallel and perpendicular to the *x* axis are defined as *s* and *p* polarizations, respectively.

The CP-dependent artifacts in our measurement setup were examined using a polycrystalline $Cu(In,Ga)Se_2$ solar cell.[50] The difference in the photocurrent between the two opposite CP states, the left CP (LCP) and right CP (RCP), was as small as $5 \pm 2 \times 10^{-4}$ relative to the photocurrent averaged over one QWP rotation (see Fig. 2S). This verifies that our measurement setup was nearly free from artifacts caused by LCP and RCP excitations. In the present study, we set the criterion that a CP-dependent photocurrent is present in a sample when the difference in the photocurrent between the LCP and RCP states is 1% or higher relative to the photocurrent averaged over one QWP rotation.

Figure 6(c) shows a schematic illustration of the model that relates the photocurrent ($I_{ph}$ in A) to the current density ($J$ in A/cm$^2$). Here, the current is generated in the excitation region in the unit of cm$^3$ and collected by a pair of electrodes adjacent to the excitation region. A photocurrent flowing in a specific direction is assumed to pass through a cross-sectional window defined by the representative width of the excitation region and the penetration depth of light, and to be collected by a pair of electrodes placed outside the windows. This is a simplified model, in that the photocurrent flowing through the window is collected without any dissipation or reduction. This model is applied in Section 6.3 to estimate the efficiency of the CPGE and a gyrotropic photoconductivity tensor.

## 6. Results and Discussion

### 6.1 Sample A

We first show zero-bias photocurrent ($I_{ph}$) data obtained from *as-cleaved* sample A: it has a quadrilateral shape with the size of around 1.5 mm × 0.8 mm, attached with a pair of electrodes separated around 0.7 mm along the $z$ axis [inset Fig. 5(a)]. The sample exhibited ohmic characteristics with resistance $R$ of around 1.6 kΩ (see Fig. 1S), and the magnitude of the $I_{ph}$ value was approximately 0.8–0.9 nA when the central part of the area $A = 1.26 \times 10^{-5}$ $cm^2$ was excited by a laser beam of 50.5 mW at normal incidence ($\theta = 0°$). The power density under these conditions was 4000 $W/cm^2$.

Tilting the device around the $x$ axis led to a polarization-dependent photocurrent. Figure 7(a) shows the $I_{ph}$ data as a function of QWP angle ($\varphi_Q$) obtained after the accumulated illumination time of $t = 2$ h with the tilt angle of $\theta = 45°$. Excitation was kept at 4000 $W/cm^2$, with the initial polarization of $p$-polarization at $\varphi_Q = 0°$. Two distinct features are observed: first, the $4\varphi_Q$ periodicity, which reflects alteration of $I_{ph}$ values between LP and CP, and second, modulation with the $2\varphi_Q$ periodicity at two different CP states, LCP and RCP, as pointed by green arrows in the figure. These facts are reminiscent of both the linear PGE (LPGE) and CPGE reported by earlier researchers.[20-22, 31-34] The $I_{ph}$ value at the LCP was higher than that at the RCP. The difference in $I_{ph}$ between the LCP and RCP excitations, $\Delta I_{ph}$, is 0.0082 nA, indicating a modulation of 3.4% relative to the average $I_{ph}$ value of 0.24 nA. It is quite encouraging that a circular polarization-dependent photocurrent is observed through the measurement across a distance on the order of millimeters, in which various inelastic carrier scattering processes are involved. A reduction in $\Delta I_{ph}$ values was observed in repeated experiments. However, some local spots retained CP-dependent $I_{ph}$, as shown in Fig. 7(b), which shows the data obtained after $t = 7$ h. Further repeating experiments resulted in a somewhat unstable $I_{ph}$ and a loss of polarization-dependent characteristics, as shown in Fig. 7(c), which was obtained after $t = 11$ h.

The observed light-induced deterioration led us to suspect that the sample underwent photoinduced oxidation and amorphization.[46] To address this problem, the light-illuminated surface of the *as-cleaved* sample A was removed by a mirror polishing process, and its polished surface was then coated with amorphous $SiO_2$ layers. The surface condition of *$SiO_2$-coated* sample A was evaluated together with *in-situ* monitoring of oxidation by Raman scattering, as described in Section 4.

The photocurrent in the regenerated *$SiO_2$-coated* sample A decreased by one order of magnitude compared to that of the fresh *as-cleaved* sample A. This suggests an increase in the density of surface recombination centers, presumably resulting from polishing and $SiO_2$ coating. Nevertheless, recovery of the polarization-dependent photocurrent was clearly observed (see Fig. 3S), as evidenced by the data obtained after an accumulated illumination time $t$ of approximately 6.5 h with an excitation power density of 4450 W/cm$^2$. The magnitude relation of the $I_{ph}$ values between the LCP and RCP is the same as that observed for the *as-cleaved* sample A.

With further continuous measurements, the fluctuations in the photocurrent gradually became pronounced, and the polarization-dependent characteristics gradually deteriorated and were completely lost after $t$ = 13 h. However, the reduction in the average photocurrent was not obvious. Considering the Raman data shown in Fig. 5(b), the observed deterioration may not be simply due to light-induced oxidation.

### 6.2 Sample B

The results of the experiments with sample A led us to an attempt to circumvent the unknown deterioration in the CP-dependent photocurrent by reducing the illumination power. We now turn our attention to the experimental results obtained from sample B, which was cut from a mother piece that was different from sample A. The lengths of the short and long sides were around 0.7 and 0.9 mm, respectively. The former is parallel to the $x$ direction, whereas the latter is parallel to the $z$ direction. Two pairs of electrodes were formed: pairs 2–3 parallel to the $x$ axis and pairs 1–4 parallel to the $z$ axis. Values of resistance are $R \approx 4.6$ and 3.1 kΩ, respectively (see Fig. 1S). A photograph of the sample is presented in Fig. 6(a). The $\varphi_Q$-dependent $I_{ph}$ data measured before the $SiO_2$ coating exhibited a clear feature of CP-dependent photocurrent in a relatively large photocurrent background. In light of the criterion set for the CP dependent photocurrent (section 5), the measured $I_{ph}$ data are regarded outside of the criterion, and are shown in Fig. 4S in the Supplementary Information.

Figure 8(a) shows the $I_{ph}$ data for the *$SiO_2$-coated* sample B, which was measured along the $z$ direction through pairs 1–4 after an accumulated illumination time of $t$ = 5.5 h. The initial polarization state at $\varphi_Q$ = 0° was linear $p$-polarization. Considering precautions against the deterioration in the polarization-dependent photocurrent, the excitation power was reduced to 2.1 mW while maintaining the same excitation area of $A = 1.26 \times 10^{-5}$ cm$^2$. Consequently, the power density decreased to 167 W/cm$^2$, and $I_{ph}$ decreased accordingly. The data obtained at normal incidence ($\theta$ = 0°) slightly show the $4\varphi$ periodicity, which is most likely due to the polarization-dependent reflection caused by the slight misalignment of the [120] axis occurring at the stage of sample mounting. Tilting the device around the $x$ axis results in enhanced $I_{ph}$ and pronounced $4\varphi_Q$ dependence, as represented by the data obtained at $\theta = 45°$. The $\varphi_Q$ dependence will be discussed later in the next section. Most importantly, modulation with the $2\varphi_Q$ periodicity is superimposed on the $4\varphi_Q$ dependence, as indicated by the green arrows in the figure. Interestingly, the magnitude relationship of $I_{ph}$ between the LCP and RCP was the same as that in sample A.

Figure 8(b) shows the $I_{ph}$ data measured along the $x$ direction through pairs 2–3, which are perpendicular to the $z$ axis. The overall magnitudes of photocurrent are small, and both $4\varphi_Q$ and $2\varphi_Q$ become less noticeable, leaving $I_{ph}$ profiles rather featureless. This indicates the presence of a strong in-plane anisotropy in the photocurrent in the $z$ and $x$ directions.

These observations lead us to the conclusion that a CP-dependent photocurrent can clearly be detected along the $z$ direction, even in handmade millimeter-sized devices whose size and shape are engineered by mechanical means. For both samples, excitation with LCP yielded a larger photocurrent than excitation with RCP, which suggests that both samples share the same chirality, probably the same as the mother pieces, right-handed Te ($P3_121$).

The contribution from the circular photon drag effect (CPDE) is remote in the present experimental configuration with the plane of light incidence on the $y$-$z$ plane. In other words, considering the crystal symmetry of Te and the incidence plane, CPDE should appear along the $z$ direction (the helical axis) with the plane of light incidence on the $x$-$z$ plane.[34,51)]

We note that the deterioration in the polarization-dependent characteristics occurred again for *$SiO_2$-coated* sample B, even under relatively weak excitation conditions. In this particular case, the photocurrent gradually decreased as the accumulated illumination time $t$ increased, and became comparable to the resolution limit of approximately 1 pA after approximately $t$ = 13 h. Before the state of complete deterioration, CP-dependent characteristics were barely noticeable, as shown in Fig. 5S for $I_{ph}$ data measured after an accumulated illumination time $t$ = 10 h with $s$-polarization. It is interesting to note that, although the phase of entire $4\varphi_Q$ modulation is reversed in the case of $s$-polarization excitation, magnitude relation between LCP and RCP states is preserved. This observation will be discussed in the next section.

Based on the Raman scattering data shown in Fig. 5, the observed deterioration in the CP-dependent $I_{ph}$ cannot be attributed to light-induced oxidation and subsequent amorphization.[46)] Taking account of the XRD characterization (Fig. 3) and possible conformations of Te,[46)] we suppose that an amorphous-like phase consisting of various forms of polymer chains exists in between the misaligned crystalline textures. These chains gradually crystallize during light illumination,[12-14)] generate trapping and/or recombination centers at the grain boundaries, and affect the nonequilibrium population of the photocarriers. We believe that the observed deterioration may be suppressed in samples with higher crystal quality and/or by using longer excitation wavelengths.

### 6.3 Discussion

The observed anisotropic zero-bias photocurrents strongly suggest that they can be attributed to the PGE involving the crystal symmetry of Te. In general, the phenomenological optical response can be represented by third-rank tensors.[42] We formulated the relationship between the optical excitation power and the resultant anisotropic photocurrent in the Te crystal by considering the crystal point group of $D_3$,[52] which resulted in a nonzero conductivity tensor:

$$\sigma_{yyx} = \sigma_{yxy} = \sigma_{xyy} = -\sigma_{xxx}$$
$$\sigma_{xyz} = -\sigma_{yxz}$$
$$\sigma_{xzy} = -\sigma_{yzx}$$
$$\sigma_{zxy} = -\sigma_{zyx}$$

Note that tensors consist of real and imaginary parts,[53] $\sigma_{ijk} = \sigma'_{ijk} - i\sigma''_{ijk}$.

Starting from the general expression of $J_i = \sigma_{ijk} E_j E_k^*$, equations of anisotropic photocurrent densities for the experimental configuration with $p$-polarization at $\varphi_Q = 0°$ are expressed by Eqs. (1) to (4). For **J** ∥ **z**, the component due to the LPGE, $J_{\mathrm{LPGE}}$, vanishes, whereas the component attributed to the CPGE, $J_{\mathrm{CPGE}}$, remains: namely,

$$J_{\mathrm{CPGE}} = -2\sigma''_{zyx} \sin\theta \sin 2\varphi_Q \quad (1)$$

$$J_{\mathrm{LPGE}} = 0 \quad (2)$$

For **J** ⊥ **z**, the relation between the LPGE and CPGE changes as follows:

$$J_{\mathrm{CPGE}} = 0 \quad (3)$$

$$J_{\mathrm{LPGE}} = \frac{\sigma'_{xyy}}{4}\{(\sin\theta)^2(\cos 4\varphi_Q + 3) + \cos 4\varphi_Q - 1\} - \frac{(\sigma'_{xzy}+\sigma'_{xyz})}{4}\sin 2\theta(\cos 4\varphi_Q + 3) \quad (4)$$

In these equations, the strengths of the fields $E_m$ and $E_n$, and thus $E_m E_n^*$ in the units of W/cm$^2$, are regarded as unity, whereas $J_i$ is the current density in the units of A/cm$^2$. Note that the $J_{\mathrm{CPGE}}$ is rather simple in the case **J** ∥ **z**, whereas the $J_{\mathrm{LPGE}}$ is rather complicated in the case of **J** ⊥ **z**. In the case of $s$-polarization, the relation between $J_{\mathrm{CPGE}}$ and $\varphi_Q$ shifts 90°, and the sign of the RHS of Eq. (1) is reversed; importantly, however, the magnitude relation between the excitation with LCP and RCP is unchanged, which can be verified through the data shown in Fig. 8(a) and Fig. 4S.

In practice, the photocurrent along the $z$ direction can be expressed phenomenologically using four terms, as expressed by Eq. (5).[54)]

$$I_{\mathrm{ph}} = C\sin 2\varphi_Q + L_1 \sin 4\varphi_Q + L_2 \cos 4\varphi_Q + D \qquad (5)$$

Here, the first and second terms correspond to the CPGE and LPGE, respectively, whereas the third and fourth terms are associated with the polarization-dependent absorption/reflection and polarization-independent terms caused by the difference in electrochemical potential between the illuminated and unilluminated regions, respectively.

Figure 9(a) shows the results of numerical fitting of the experimental $I_{\mathrm{ph}}$ data shown in Fig. 8(a). The fitting yielded $C \sim 0.18$ pA, $L_1 \sim -0.18$ pA, $L_2 \sim 0.98$ pA, and $D \sim 13$ pA, with significant figures of two. We notice a relatively large contribution of the $L_2$ and $D$ terms, whereas the contributions of the $C$ and $L_1$ terms, both being PGE terms, are relatively small. A breakdown of the numerical fitting for each $\varphi_Q$-dependent term in Eq. (5), which facilitates the determination of the contribution of each term, is shown in Fig. 9 (b). The most notable $L_2$ term can qualitatively be explained in terms of LP-dependent absorption of crystalline Te: the absorption coefficient is larger for light polarization parallel to the $z$ axis (**E** ∥ **z**) than that for **E** ⊥ **z**.[3)] Thus, photocurrent is largest when $p$-polarization light is impinged on Te at $\varphi_Q = 90° n$ $(n = 0, 1, 2, \ldots)$, whereas it is smallest when CP light is impinged at $\varphi_Q = 45°\,(2n + 1)$ ($n$ = 0, 1, 2, ...), where the $p$-polarization component is reduced by half. Polarization-dependent reflection[55)] also helps enhance the difference in the photocurrent between the two polarization states. Within the context of a polarization-dependent absorption scenario, the magnitude relation in photocurrent between $\varphi_Q = 90° n$ and $\varphi_Q = 45°\,(2n + 1)$ is inverted when $s$-polarization is used for excitation (Fig. 4S). Therefore, $L_2$ term in Eq. (5) can be classified as a polarization-dependent first-order effect term associated with light absorption.

Let us now examine the validity of nonzero $L_1$ and $C$ coefficients. These values are smaller than the $L_2$ value, reflecting higher-order phenomena (Eqs. 1 and 4). A careful inspection of Fig. 9(b) reveals the unique contributions of $L_1$ and $C$ terms. First, the contribution of the $L_1$ term manifests as a slight asymmetrical distortion superimposed on the cosine waveform of the $L_2$ term. This suggests that small fluctuations in the photocurrent may result in a nonzero $L_1$ coefficient, and one must be careful when dealing with small $L_1$ values. It was difficult to determine the contribution of the $L_1$ term solely from the experimental data obtained in the present study.

In contrast, the contribution of the $C$ term is rather straightforward: it manifests itself as a periodic amplitude modulation at $\varphi_Q = 45°\,(2n + 1)$, which is relatively easy to read from experimental data when background signals are reasonably stable. We can recognize these points in Figs. 6S(a) and (b), in which calculations without the $L_1$ and $C$ terms are shown, respectively. Consequently, we may be able to state that validity of the $C$ coefficient is relatively higher than that of the $L_1$ coefficient.

Table 1 lists the values of the four coefficients obtained by fitting the experimental data to Eq. (5). Clearly, the polarization-independent $D$ term decreases with both photodeterioration and surface polishing. Interestingly, the ratio $C/D$ remains nearly constant, 0.01–0.02, for both samples A and B, no matter excitation is relatively weak or strong. This fact suggests that CPGE and nonpolarization-dependent photocurrent driven by the electrochemical potential gradient may be correlated. The range of the ratio $L_2/D$ is a little wide, 0.04–0.09, for which the difference in reflection due to surface roughness may be responsible. In case of excitation with *s*-polarization, polarities of $C$ and $L_2$ are reversed, as anticipated from Eq. 1.

Analyses of the photocurrent data obtained along the $x$ direction at $\theta = 0°$ (normal incidence) were also carried out using the data shown in Fig. 8(b). At $\theta = 0°$, Eq. (4) is simplified as $I_{ph} - D = J\,(\perp \mathbf{z}) \cdot A' = L_1'(\cos 4\,\varphi_Q - 1)$, where $L_1' = \frac{\sigma'_{xyy} \cdot A'}{4}$ with $A'$ the area of cross section through which a current flows [Fig. 6(c)]. The quantity $I_{ph} - D$ can be obtained from the difference between the maximum and minimum photocurrents, $I_{max}$ and $I_{min}$, at $\varphi_Q = 90°n, (n = 0, 1,\ ..)$ and $\varphi_Q = 45°(n+1), (n = 1,\ 3,\ \dots)$, respectively. The average values are $I_{max}$ = 4.63 pA and $I_{min}$ = 4.57 pA, from which $L_1'$ of 0.06 pA is estimated. This suggests that the effect of LPGE on Te was not significant.

Based on the $C$ value extracted from the data shown in Fig. 8(a) for *$SiO_2$-coated* sample B, the efficiency of the CPGE is estimated to be around $1.1 \times 10^{-15}$ A·cm$^2$/W with an excitation power density of 167 W/cm$^2$. With this value, we are able to estimate the value of $2\sigma''_{zyx}$ in Eq. (1), using the simplified model presented in Sections 5 and Fig. 6(c). Note that $2\sigma''_{zyx}$ is tied together with the gyrotropic conductivity tensor $\beta$ through $\beta_{ij} = \varepsilon_{jkl}\mathrm{Im}\{\sigma_{ikl}\} = -\varepsilon_{jkl}\sigma''_{ikl}$ in which $\beta_{ij}$ is the second-rank axial tensor, $\varepsilon_{ijk}$ the third-rank Levi-Civita tensor, and $\sigma_{ikl}$ the third-rank polar tensor.[44)] The area of the cross-sectional window $A'$ is estimated by the diameter of the excitation spot, $\phi$, multiplied by the penetration depth of light, $d$, that is, $A'$ = 40 μm × 50 nm = $2 \times 10^{-8}$ cm$^2$. Thus, $2\sigma''_{zyx} = \frac{1\times10^{-15}\ \mathrm{A{\cdot}cm^2/W}}{2\times10^{-8}\ \mathrm{cm^2}} = 5 \times 10^{-8}\ \mathrm{A/W}$ = 50 nA/W, with one significant figure. With respect to data obtained from other cases, $\beta$ value is around 45 nA/W for the as-cleaved sample A with an excitation power of 4000 W/cm$^2$, whereas it is around 5 nA/W for the $SiO_2$-coated sample A with an excitation power of 4450 W/cm$^2$. The $\beta$ value obtained for the as-cleaved sample B is disclosed in Supplementary Information (850 nA/W).

Depending on the sample conditions, the extrinsic $\beta$ values range from a few nA/W to a few tens of nA/W or may even extend up to a few hundreds of nA/W. Tentatively, we are able to notice a trend: the $\beta$ value is larger for the as-cleaved samples than that for the $SiO_2$-coated samples. It is also interesting to note that the $\beta$ values extracted in the present study are greater than those reported for intraband excitation in bulk Te (3.2 nA/W).[34)] Studies with single-crystal Te samples should be pursued to further reveal the difference between intraband and interband excitations.

Finally, we review the magnitude of the CPGE obtained earlier with Te flakes within the context of our framework. Experiments with Te flakes of a few tens-μm using a CW $CO_2$ laser[31)] showed CPGE of around 1.4 nA at 680 W/cm$^2$ with excitation at wavelength $\lambda$ = 10.6 μm, which corresponds to the efficiency of $1.8 \times 10^{-12}$ A·cm$^2$/W and extrinsic $\beta \sim 180$ μA/W. The absence of CPGE under excitations at $\lambda$ = 4.0 and 4.5 μm can be understood in terms of the direct band-to-band transitions around the *H* point, where spin states cross over and spin-dependent optical transitions are relaxed (Fig. 2).

Experiments with Te flakes of size around 100 μm excited by a CW-laser at $\lambda$ = 633 nm up to the power of 0.6 mW showed CP-dependent photocurrent of 10.9 nA/mW,[32)] which corresponds to the efficiency of $4.1 \times 10^{-12}$ A·cm$^2$/W and extrinsic $\beta \sim 1200$ μA/W referring to the diameter of an excitation spot of around 7 μm.

The above-mentioned estimations suggest two interesting observations: firstly, the values of $\beta$ are larger in the Te flakes than those in bulk Te, and secondly, the $\beta$ value is larger for interband excitation than for intraband excitation, also in Te flakes.

CPGE takes place when optical excitation breaks the balance of the carrier population between the two opposite spins in *K*-space. For intraband excitation, this imbalance occurs without a change in carrier numbers (Fig. 2). In contrast, interband excitation yields excess spin-polarized carriers which relax primarily via optical phonon emissions and merge in the carrier bath. Consequently, the number of spin-polarized carriers increases within the limit determined by the lifetime of excess carriers, and the imbalance in the carrier population between the two opposite spins is enhanced compared to the intraband excitation. This inference assumes that the energy relaxation of hot carriers is not accompanied by the spin flip event, owing to the spin-separated energy bands.

As to the difference between Te flakes and bulk Te crystals in terms of the magnitude of the $\beta$ value, physical origins remain open, particularly regarding the role of quantum size effects on band structures and enhanced surface contributions, incorporating the issue concerning crystal quality of samples, namely, single or multiple domains. We only state here that further systematic investigations are needed to fill the gap between the flake and bulk.

## 7. Conclusions

We have reported the detection of zero-bias photocurrents induced by interband excitation using oblique near-infrared light ($\lambda$ = 0.780 μm) along the directions parallel and perpendicular to the helical axis in millimeter-sized Te samples. A CP-dependent photocurrent is observed exclusively along the direction parallel to the helical ($z$) axis. The unforeseen deterioration of polarization-dependent characteristics with an increase in accumulated illumination time has also been reported. A framework for estimating the gyrotropic photoconductivity tensor $\beta$, which is the ratio between the circular-polarization-dependent photocurrent density and light intensity, has been developed, through which the extrinsic $\beta$ values have been found to span from a few nA/W to a few tens of nA/W, depending on the sample conditions. Interestingly, these values appear greater than those reported for the intraband excitations in bulk Te (around 3.2 nA/W).[34)] The developed framework is also applied to experimental data obtained from Te flakes,[31,32)] which also suggests that the extrinsic $\beta$ value with the interband excitation is larger than that with the intraband excitation.

## Acknowledgements

We gratefully acknowledge the Materials Analysis Division and the Semiconductor and MEMS Processing Division, both in the Core Facility Center at the Institute of Science Tokyo, for X-ray diffraction measurements and RF sputtering of $SiO_2$ layers, respectively. We thank A. Iwase and Y.-C. Huang, student members at the Satoh laboratory, for theoretical discussions and technical assistance. This work is supported in part by JSPS KAKENHI (Grant Nos. JP21H01032 and JP22H01154), MEXT X-NICS (Grant No. JPJ011438), NINS OML Project (Grant No. OML012301), and JST CREST (Grant No. JPMJCR24R5).

## Supplementary Information

The posted online data are experimental data related to current-voltage data of samples A and B (Fig. 1S), the validation of the photocurrent measurement setup (Fig. 2S), photocurrent data obtained from regenerated sample A (Fig. 3S), as-cleaved sample B (Fig. 4S), $SiO_2$-coated sample B before the stage of complete deterioration (Fig. 5S), and inverse analysis with incomplete phenomenological equations (Fig. 6S).

## References

1) A. Koma, E. Takimoto, and S. Tanaka, Phys. Stat. Solidi **40**, 239 (1970).

2) K. C. Nomura, Phys. Rev. Lett. **5**, 500 (1960).

3) S. Tutihasi, G. G. Roberts, R. C. Keezer, and R. E. Drews, Phys. Rev. **177**, 1143 (1969).

4) N. Miura and S. Tanaka, phys. stat. sol. **42**, 257 (1970).

5) N. Miura, H. Shinno, and S. Tanaka, Solid State Commun. **9**, 971 (1971).

6) R. Yoshizaki and S. Tanaka, J. Phys. Soc. Jpn. **30**, 1389 (1971).

7) L. Rothkirch, R. Link, W. Sauer, and F. Manglus, phys. stat sol. **31**, 147 (1969).

8) A. Nussbaum and R.J. Hager, Phys. Rev. **123**, 1958 (1960).

9) C.H. Chanpness and A.L. Kipping, J. Can. Phys. 48, 3038 (1970).

10) P. L. Weimer, Proc. IEEE **52**, 608 (1964).

11) H. L. Wilson and W. A. Gutierrez, Proc. IEEE Lett. **55**, 415 (1967).

12) J. Feinleib, J. deNeufville, S. C. Moss, and S. R. Ovshinsky, Appl. Phys. Lett. **18**, 254 (1971).

13) A. E. Bell and F. W. Spong, Appl. Phys. Lett. **38**, 920 (1981).

14) P. C. Clemens, Appl. Opt. **22**, 3165 (1983).

15) J. R. Reitz, Phys. Rev. **105**, 1233 (1957).

16) I. A. Firsov, Soviet Phys. JETP **5**, 1101 (1957).

17) T. Doi, K. Nakao, and H. Kamimura, J. Phys. Soc. Jpn. **28**, 36 (1970).

18) T. Doi, K. Nakao, and H. Kamimura, J. Phys. Soc. Jpn. **28**, 822 (1970).

19) K. Nakao, T. Doi, and H. Kamimura, J. Phys. Soc. Jpn. **30**, 1400 (1971).

20) E. L. Ivchenko and G. E. Pikus, JETP Lett. **27**, 604 (1978).

21) V. M. Asnin, A. A. Bakun, A. M. Danishevskii, E. L. Ivchenko, G. E. Pikus, and A. A. Rogachev, JETP Lett. **28**, 74 (1979).

22) V. M. Asnin, A. A. Bakun, A. M. Danishevskii, E. L. Ivchenko, G. E. Pikus, and A. A. Rogachev, Solid State Commun. **30**, 565 (1979).

23) D.Q. Yang, L.Q. Zhu, J.L. Wang, W. Xia, J.Z. Zhang, K. Jiang, L.Y. Shang, Y.W. Li, and Z.G. Hu, phys. stat. sol. B **259**, 2100625 (2022).

24) X. Wan, A. M. Turner, A. Vishwanath, and S. Y. Savrasov, Phys. Rev. B **83**, 205101 (2011).

25) M. Hirayama, R. Okugawa, S. Ishibashi, S. Murakami, and T. Miyake, Phys. Rev. Lett. **114,** 206401 (2015).

26) K. Nakazawa, T. Yamaguchi, and A. Yamakage, Phys. Rev. Materials **8**, L091601 (2024).

27) T. Furukawa, Y. Shimokawa, K. Kobayashi, and T. Itou, Nat. Commun. **8**, 954 (2017).

28) T. Furukawa, Y. Watanabe, N. Ogasawara, K. Kobayashi, and T. Itou, Phys. Rev. Research **3**, 023111 (2021).

29) G. L. J. A. Rikken and N. Avarvari, Phys. Rev. B **99**, 245153 (2019).

30) S. S. Tsirkin, P. A. Puente, and I. Souza, Phys. Rev. B **97**, 035158 (2018).

31) J. Ma, B. Cheng, L. Li, Z. Fan, H. Mu, J. Lai, X. Song, D. Yang, J. Cheng, Z. Wang, C. Zeng, and D. Sun, Nat. Commun. **13**, 5425 (2022).

32) C. Niu, S. Huang, N. Ghosh, P. Tan, M. Wang, W. Wu, X. Xu, and P. D. Ye, Nano Lett. **23**, 3599 (2023).

33) E. Mönch, M.D. Moldavskaya, L.E. Golub, V.V. Bel'kov, J. Wunderlich, D. Weiss, J.V. Gumenjuk-Sichevska, C. Niu, P.D. Ye, and S.D. Ganichev, Nano Lett. **25**, 476 (2024).

34) M. D. Moldavskaya, L. E. Golub, S. N. Danilov, V. V. Bel'kov, D. Weiss, and S. D. Ganichev, Phys. Rev. B **108**, 235209 (2023).

35) T. Satoh, S.-J. Cho, R. Iida, T. Shimura, K. Kuroda, H. Ueda, Y. Ueda, B. A. Ivanov, F. Nori, and M. Fiebig, Phys. Rev. Lett. **105**, 077402 (2010).

36) T. Satoh, Y. Terui, R. Moriya, B. A. Ivanov, K. Ando, E. Saitoh, T. Shimura, and K. Kuroda, Nat. Photon. **6**, 662 (2012).

37) W.-H. Hsu, K. Shen, Y. Fujii, A. Koreeda, and T. Satoh, Phys. Rev. B **102**, 174432 (2020).

38) K. Ishito, H. Mao, K. Kobayashi, Y. Kousaka, Y. Togawa, H. Kusunose, J. Kishine, and T. Satoh, Chirality **35**, 338 (2023).

39) K. Ishito, H. Mao, Y. Kousaka, Y. Togawa, S. Iwasaki, T. Zhang, S. Murakami, J. Kishine, and T. Satoh, Nat. Phys. **19**, 35 (2023).

40) R. C. Roca, N. Nishizawa, K. Nishibayashi, and H. Munekata, J. Appl. Phys. **123**, 213903 (2018).

41) N. Nishizawa, K. Nishibayashi, and H. Munekata, Proc. Natl. Acad. Sci. U.S.A. **114**, 1783 (2017).

42) It has been defined as *photovoltaic* tensor in B. I. Sturman and V. M. Fridkin, *The Photovoltaic and Photorefractive Effects in Noncentrosymmetric Materials*, Chapter 1 (Translated from the Russian by J.E.S. Bradley, Gordon and Breach Science Publishers S. A., 1992).

43) Irradiation area on a sample is estimated based on nominal beam diameter of 0.3 mm, with distance between a collimator and a sample of 100 mm and beam dispersion angle of 0.04°. Range of Bragg's angle is determined by a two-dimensional x-ray detector with solid angle of 26°. Distance between a sample and a detector is 200 mm.

44) J. S. Blakemore, J. W. Schultz, and K. C. Nomura, J. Appl. Phys. **31**, 2226 (1960).

45) Intrinsic carrier concentration ($n_i = n = p$) at room temperature is estimated to be around $1.5 \times 10^{15}$ cm$^{-3}$ on the basis of effective density-of-state approximation with effective mass of electron[5] and hole,[6] $m_e = 0.047$ and $m_h = 0.11$, respectively, and $E_g = 0.33$ eV.[23] The Fermi level can be deduced by $E_F = \frac{E_g}{2} + \frac{3}{4} kT \cdot log_e \left(\frac{m_h}{m_e}\right) = \frac{E_g}{2} + 0.0165$ eV. For heavily p-type, the Fermi level at room temperature would be between the valence band top and the middle of the band gap.

46) T. Vasileiadis and S. Yannopoulos, J. Appl. Phys. **116**, 103510 (2014).

47) M.A. Grasser, T. Pietsch, E. Brunner, T. Doert, and M. Ruck, ChemistryOpen **10**, 117 (2021).

48) Coating of an amorphous $SiO_2$ film was carried out without intentionally heating a sample holder. Power of RF sputtering was 100W with Ar atmosphere pressure of 1.0 Pa. Deposition rate was around 7.9 nm/min. This deposition condition yields optically transparent films with the optical index $n = 1.46$.

49) A. S. Pine and G. Dresselhaus, Phys. Rev. B **4**, 356 (1971).

50) S. Ji, T. Hayakawa, N. Suyama, K. Nakada, and A. Yamada, Jpn. J. Appl. Phys. **59**, 041003 (2020).

51) V.A. Shalygin, M.D. Moldavskaya, S.N. Danilov, I.I. Farbshtein, L.E. Golub, Phys. Rev. B **93**, 045207 (2016).

52) R. W. Boyd, *Nonlinear Optics,* p. 46, Table 1.5.2 (4th Ed., Academic Press, 2020).

53) F. Hennerberger, N. S. Averkiev, and R. J. Rasulov, phys. stat. sol. (b) **109**, 343 (1982).

54) J. W. McIver, D. Hsieh, H. Steinberg, P. Jarillo-Herrero, and N. Gedik, Nat. Nanotechnol. **7**, 96 (2012).

55) Polarization-dependent reflectivity $R$ at $\theta = 45°$ is estimated to be around 0.40 and 0.61 for $\mathbf{E} \parallel \mathbf{z}$ and $\mathbf{E} \perp \mathbf{z}$ at around 1.5 eV, respectively, on basis of dielectric constants of $\varepsilon_1 \sim$ 40 and 34, respectively, extracted by the work the Ref. 3.

# Figures

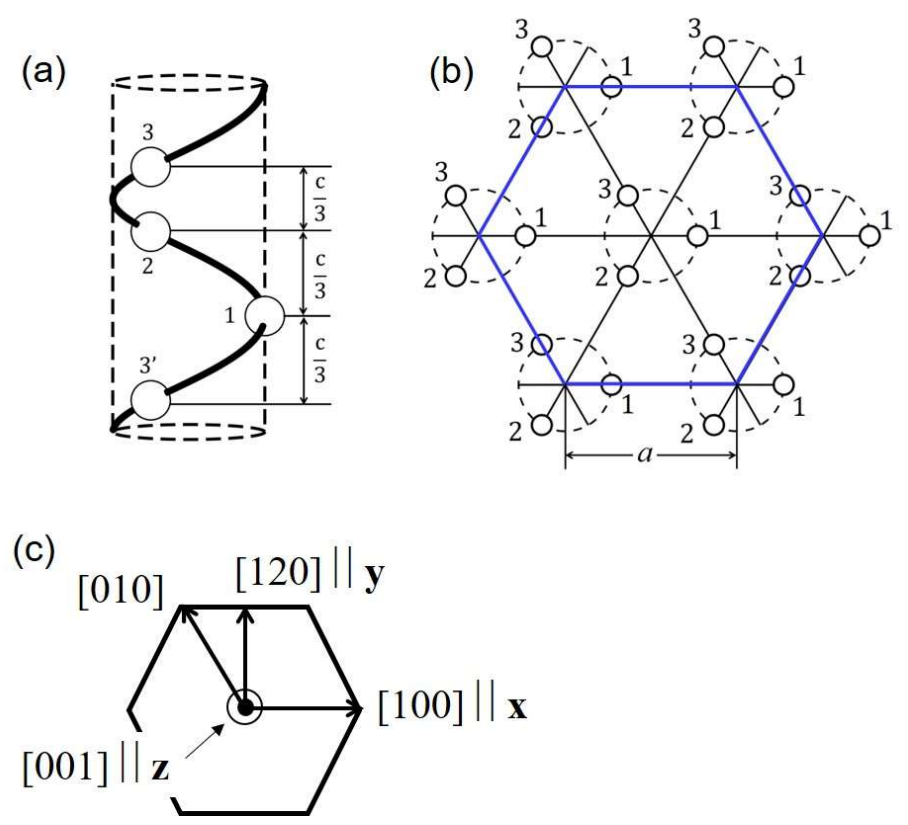

Figure 1: Schematic illustrations of (a) a part of a right-handed, trigonal wire made of Te atoms, and (b) a cross-sectional view of trigonal wires packed into a hexagonal array. These figures are prepared referring to Ref. 1 with permission from Wiley. (c) Definition of crystallographic axes used in this work. Three primitive axes in a hexagonal crystal are denoted by [100], [010], and [001] in terms of Miller indices, whereas a set $(x, y, z)$ represents the orthogonal coordinate system used in the present study. Note: the [120] direction is parallel to the $y$-axis.

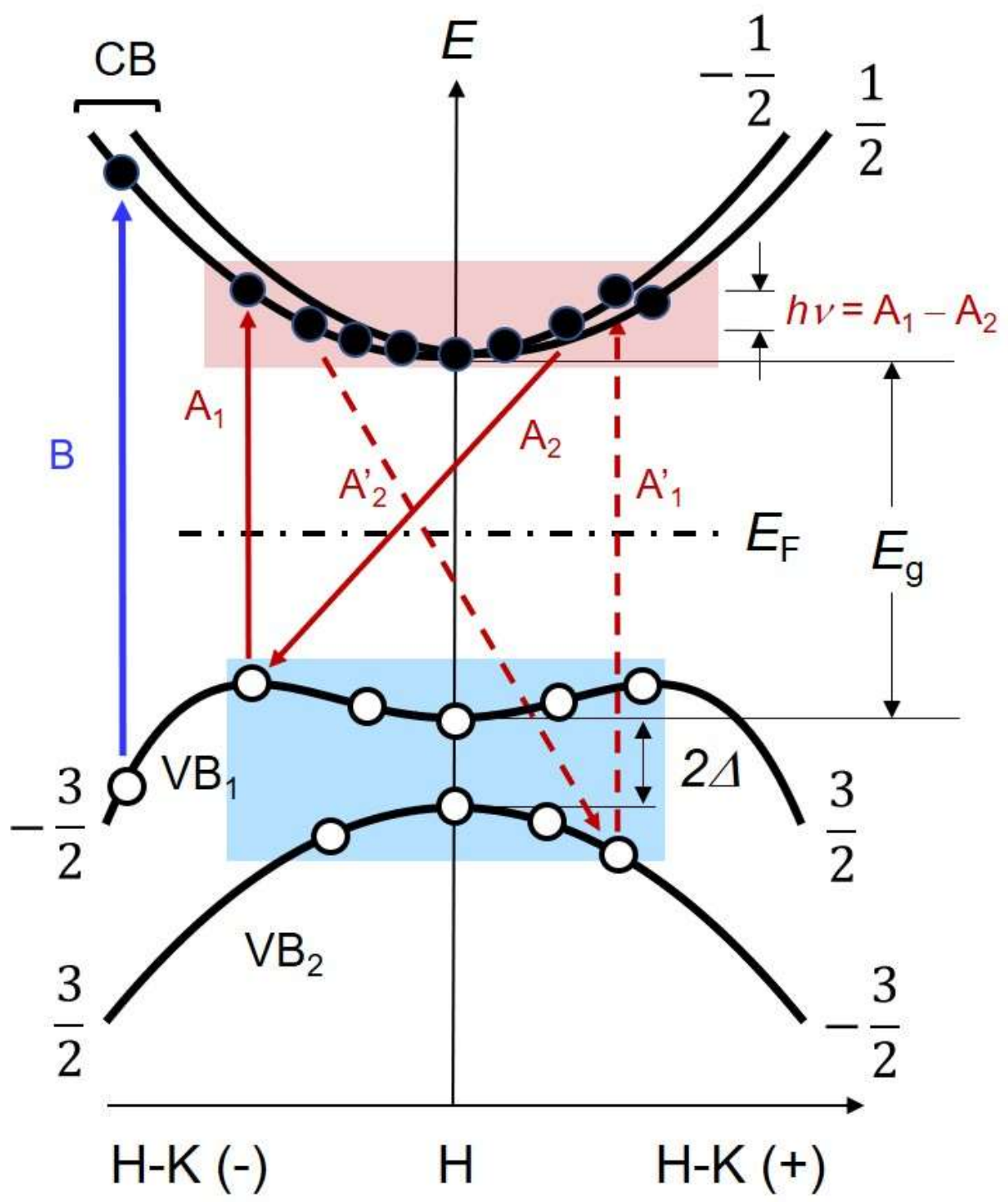

Figure 2: Schematic illustration of band structure around *H* point in *E* - *K* space, and the mechanism of CPGE considered in the landmark studies.[20-22] CB, $VB_1$, and $VB_2$ depict conduction band, the topmost valence band, and second lower valence band, respectively. Spin separation due to the lack of space inversion symmetry and spin-orbit interaction is added to the CB. Furthermore, $E_g$ and $2\Delta$ represent the band gap and energy separation between $VB_1$ and $VB_2$, respectively. According to the recent studies with infrared absorption spectroscopy,[23] $E_g$ = 0.333 eV and $2\Delta$ = 0.087 eV at 5 K, and $E_g \approx$ 0.33 eV at room temperature. Black and white dots represent coexisting electron and hole carriers, whereas $E_F$ represents the Fermi level at room temperature. A pair of transitions $A_1$ ($A'_1$) and $A_2$ ($A'_2$), starting from $VB_1$ ($VB_2$), represents a virtual intraband transition of electrons in the light-brown region with photon energy $h\nu = A_1 - A_2$. The difference in the transition probability between $A_1$ ($A_2$) and $A'_1$ ($A'_2$) should be noted. Additionally, intraband transition of holes in the light-blue region has been considered in ref. 22. The interband transition is shown by a blue arrow B. All optical transitions shown here are accompanied by $\Delta s$ =+1, namely, absorption and emission of circular polarized light $\sigma^+$.

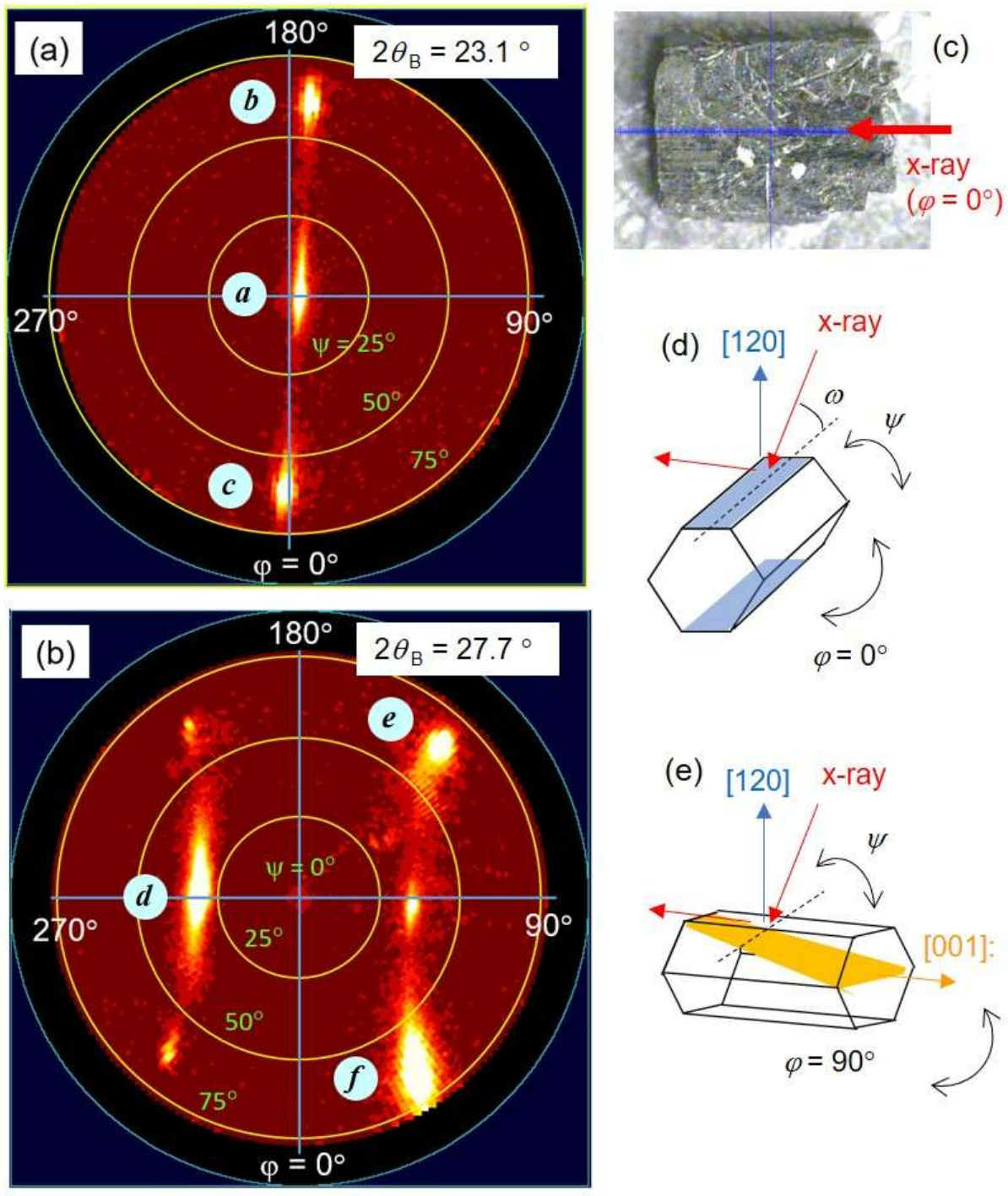

Figure 3: Pole figures obtained at two different Bragg's angles with center values $2\theta_B$: (a) 23.1° and (b) 27.7°. Three auxiliary figures depicted on the right sides of pole figures are (c) a top view picture of a mother piece added with a red arrow which indicates incidence direction of an X-ray beam at $\varphi = 0°$, (d) a hexagonal prism with the (010) plane and the [120] direction in pale blue, and (e) a hexagonal prism with the (011) plane and the [001] direction in orange. Symbols $\varphi$ and $\psi$ represent in-plane rotation and tilt angle of a sample stage relative to the incident X-ray plane. Symbols "*a*" to "*f*" on pole figures represent six different reflection planes.

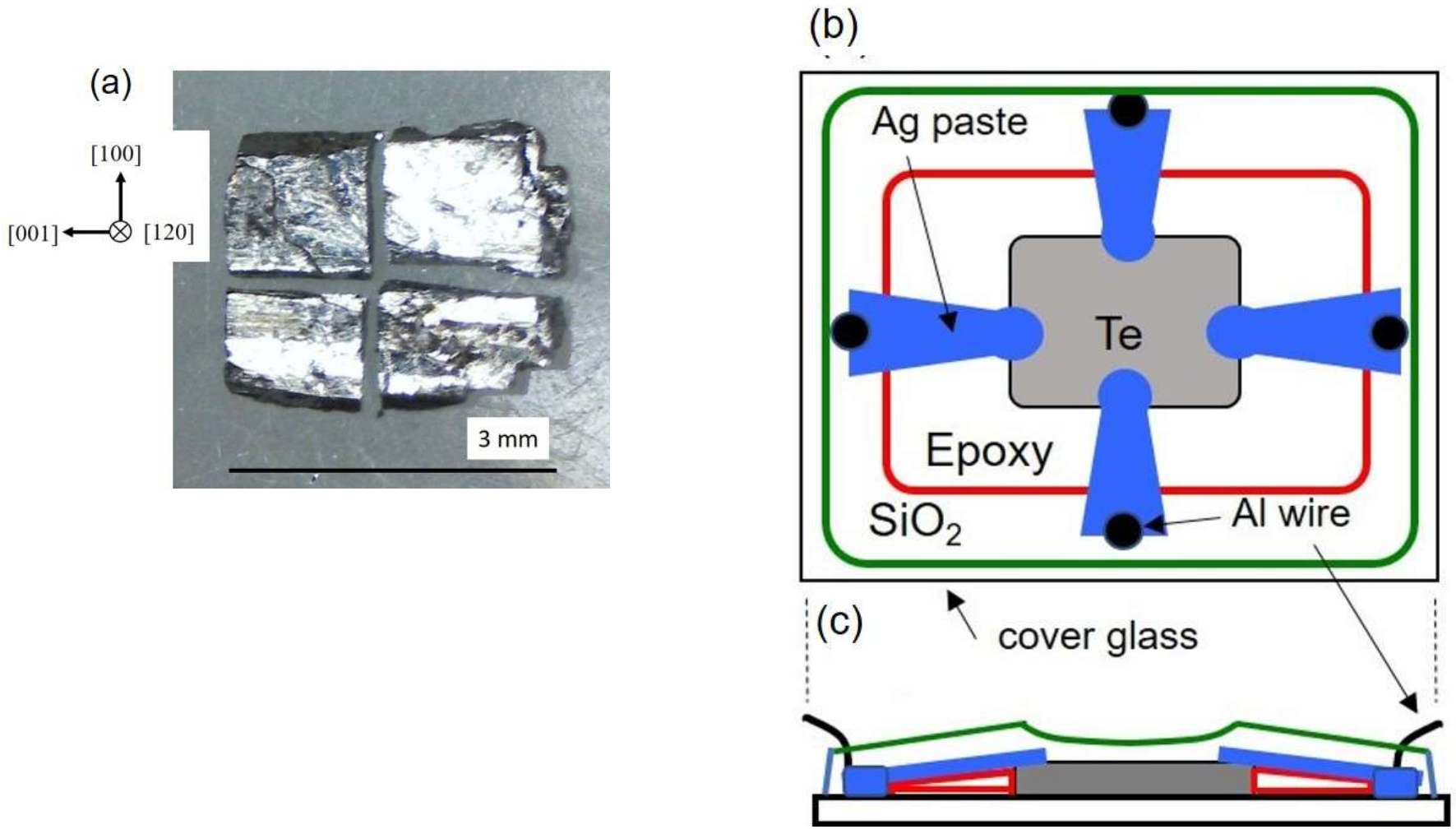

Figure 4: (a) A top-view picture of a mother piece, which is cut into four quadrilateral-shaped chips. Crystallographic axes of hexagonal Te are shown on the upper left. Schematic illustrations of (b) top and (c) cross-sectional views of a quadrilateral-shaped device, incorporating various constituent materials in the device.

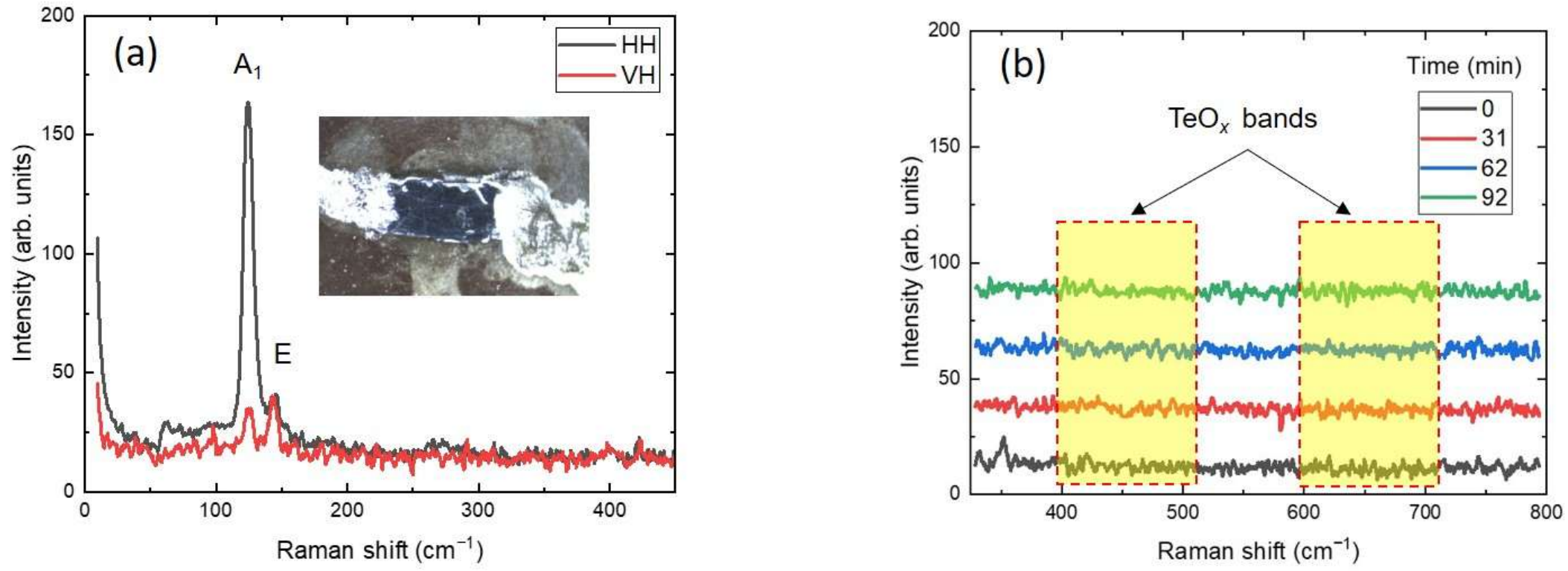

Figure 5: Raman spectra in (a) relatively low and (b) relatively high wavenumber regions. In panel (a), polarization states of backscattered and impinging light are expressed by the notation XX in which H and V denote horizontal- and vertical-linear polarization, respectively. The inset shows a top-view picture of $SiO_2$-coated sample A on which a pair of electrodes is formed parallel to the in-plane *z* direction. (b) Spectra measured with HH condition at four different accumulated illumination times. Yellow hatches represent wavenumber regions in which broad $TeO_2$ bands have been reported.[46)]

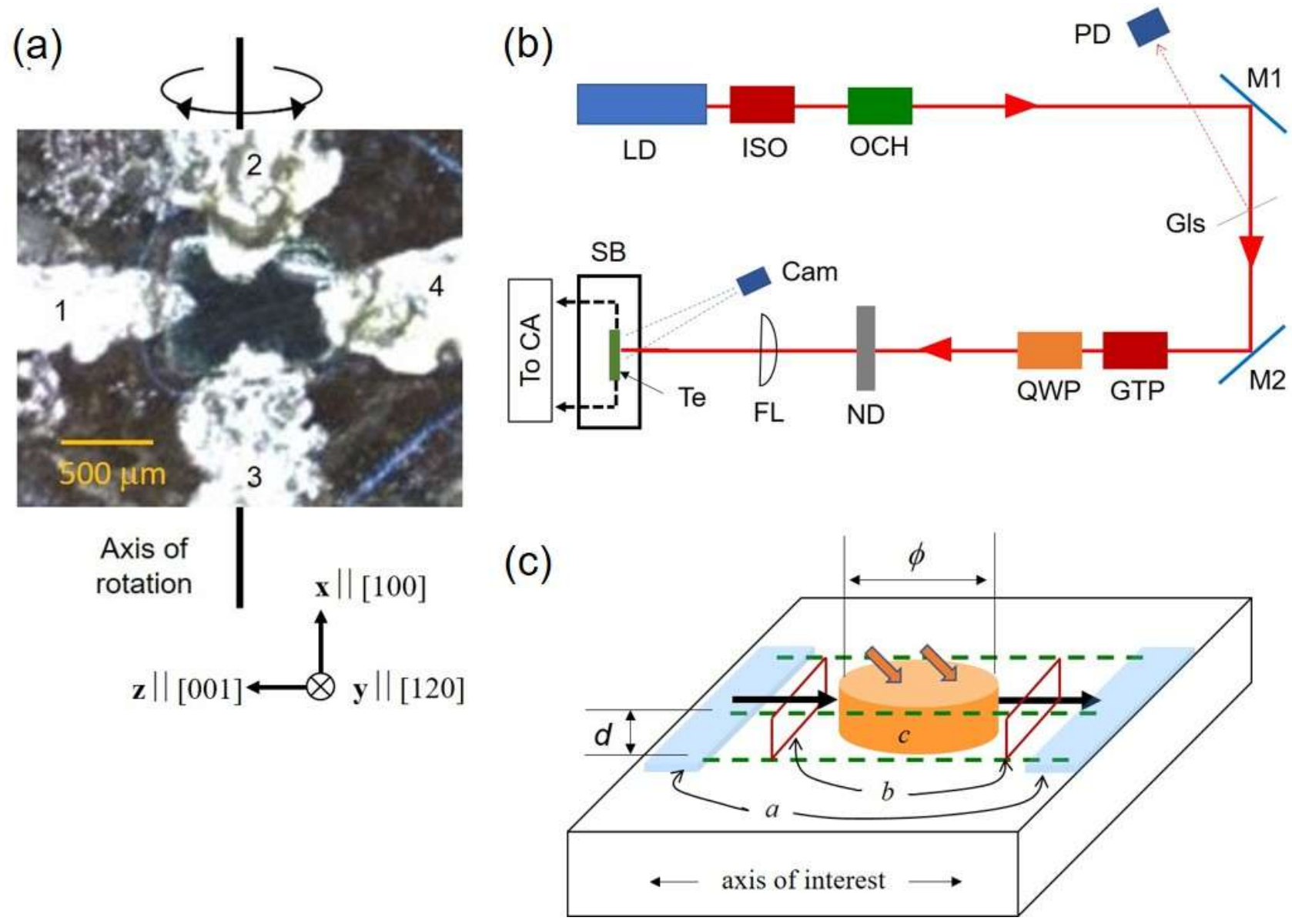

Figure 6: (a) A top view picture of *SiO₂-coated* sample B, together with an illustration which schematically shows the relationship between **x**, **y**, and **z** axes and crystal axes expressed by the Miller indices. Note that [120] direction is parallel to the axis normal to a cleaved plane. (b) Schematic illustration of the setup for polarization-dependent photocurrent measurement. LD: laser diode, ISO: isolator, OCH: optical chopper, PD: photo diode, M1 and M2: gold coated mirrors, Gls: glass, GTP: Gran-Tayler prism, QWP: quarter wave plate, ND: neutral density filter (13%), FL: focus lens (L = 70 or 100 mm), Cam: IR camera, Te: Te sample, SB: aluminum sample box, and CA: current amplifier. (c) A model illustrating the estimation of photocurrent density. Labels *a*, *b*, and *c* represent a pair of electrodes, rectangle windows through which a current flows, and an excitation region, respectively. Notations $\phi$ and $d$ denote the diameter of the excitation spot and penetration depth of light, respectively. Black arrows represent a photocurrent flowing along the axis of interest.

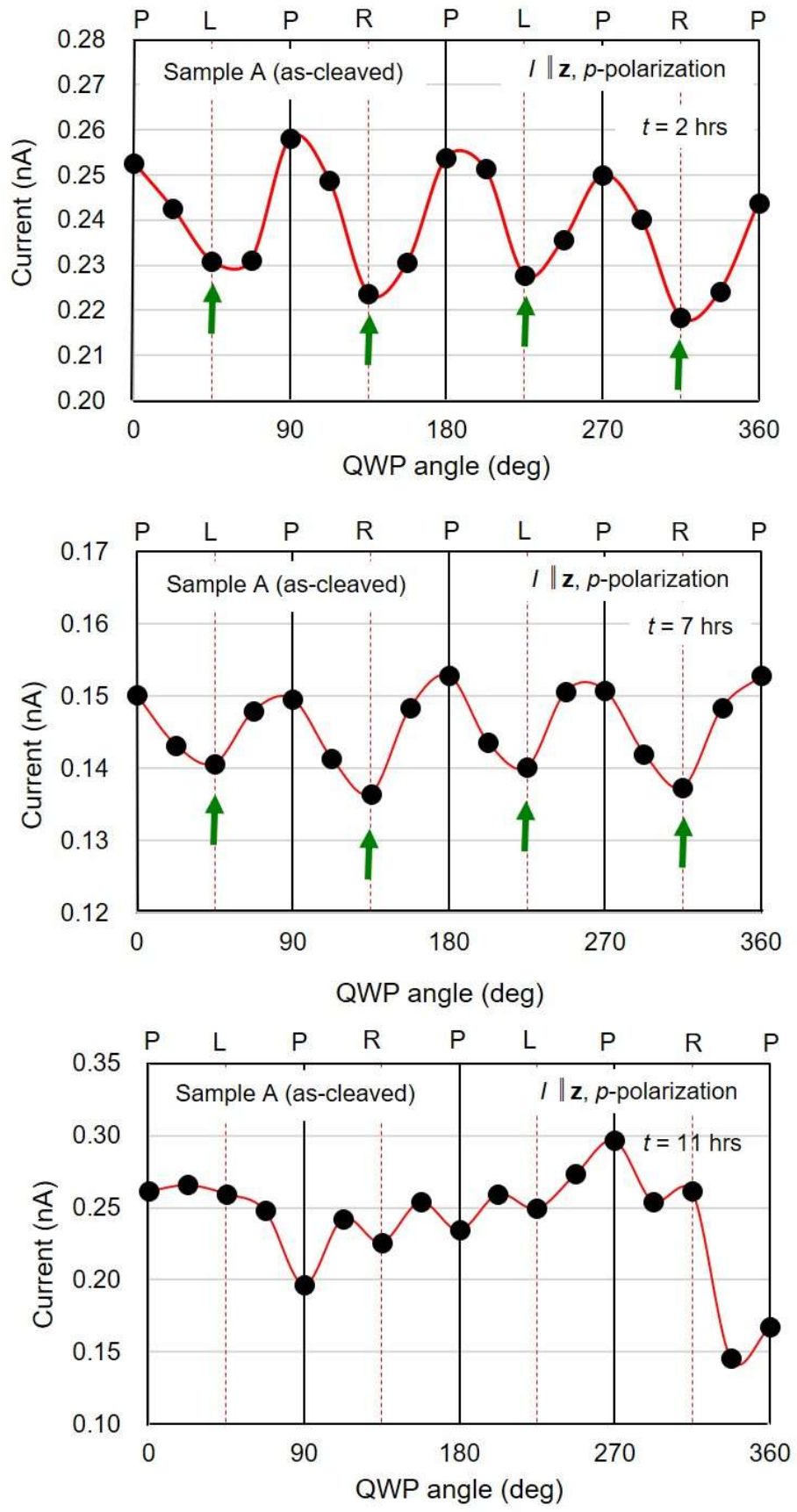

Figure 7: Photocurrent data obtained from as-cleaved sample A as a function of QWP angle ($\varphi_Q$) measured after three different accumulated illumination times, $t$ = (a) 2, (b) 7, and (c) 11 h. A pair of electrodes is formed parallel to the $z$ direction, as shown in the inset of Fig. 4 (a). Incident angle of light is $\theta = 45°$ with $p$-polarization at $\varphi_Q = 0°$, and excitation power density is 4000 W/cm$^2$. Notations P, L, and R indicate $p$, left-circular, and right-circular polarizations, respectively. Red curved lines are for eye guide.

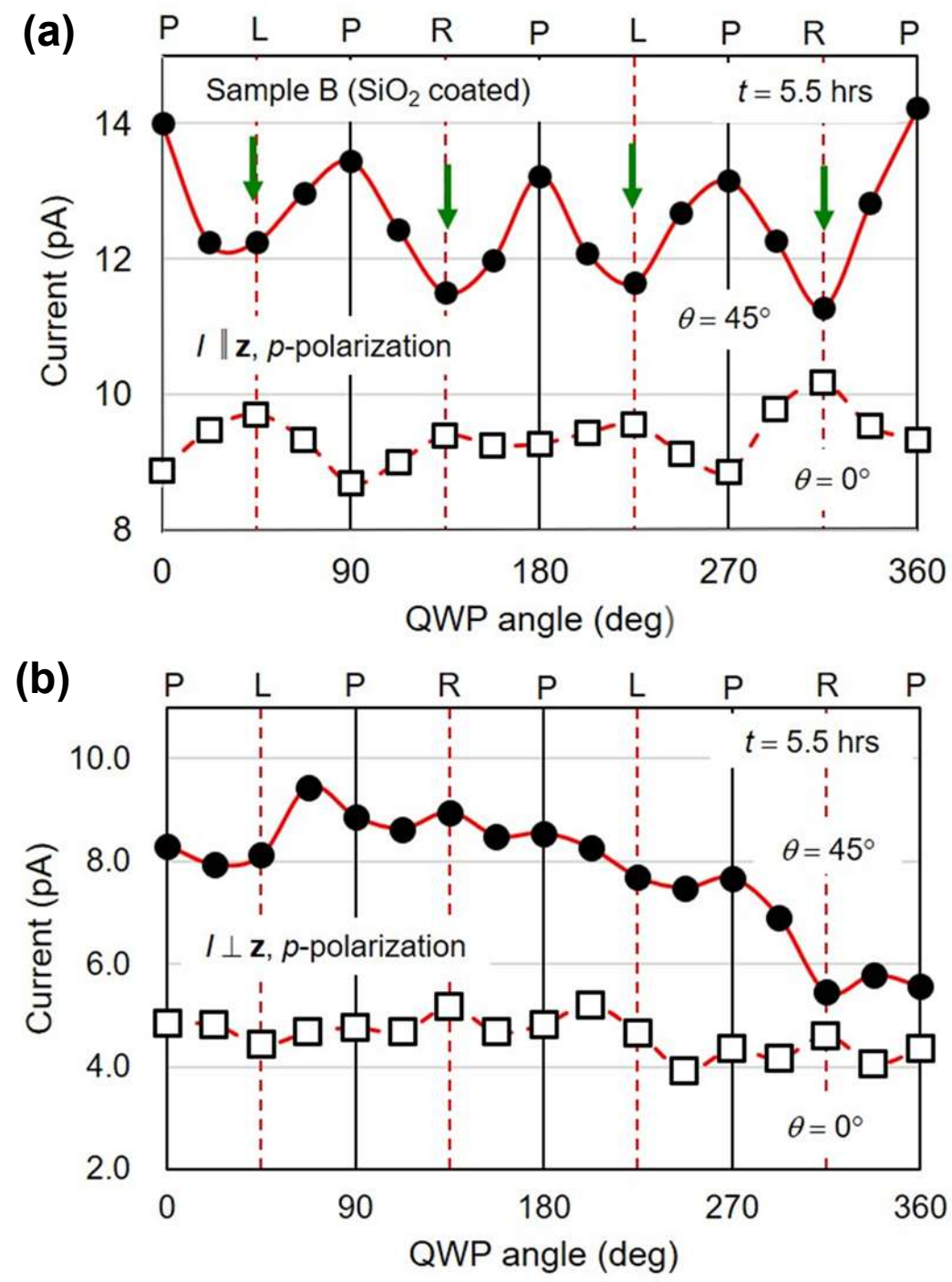

Figure 8: (a) Photocurrent data obtained from *$SiO_2$-coated* sample B as a function of rotation angle of QWP ($\varphi_Q$). Measurements were performed after a total illumination time of $t$ = 5.5 h at two different light-incident angles $\theta$= 0 and 45°, with pairs of electrodes (a) parallel and (b) perpendicular to the $z$ direction. Initial polarization at $\varphi_Q = 0°$ is $p$-polarizations with excitation power density 167 W/cm$^2$. Notations P, L, and R indicate $p$, left-circular, and right-circular polarizations, respectively. Red curved lines are for eye guide.

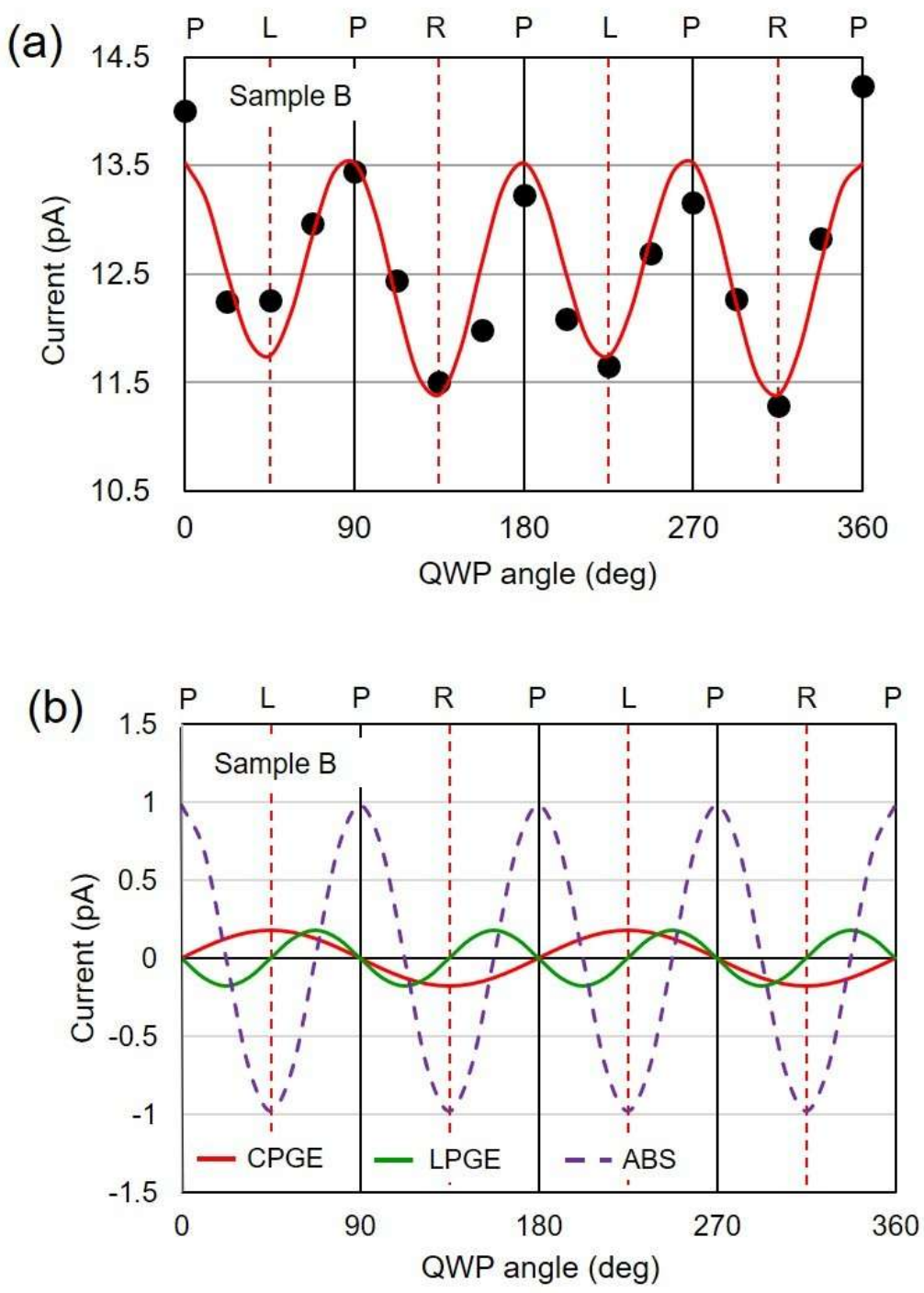

Figure 9: (a) A photocurrent curve in red calculated with Eq. (5) based on experimental data (black dots) shown in Fig. 7(a) with tilt angle $\theta$ = 45°. (b) Breakdown of a calculated photocurrent: CPGE component in a solid red line, LPGE component in a solid green line, and polarization-dependent absorption component in a dashed purple line. Notations P, L, and R indicate $p$, left-circular, and right-circular polarizations, respectively.

Table 1: A list of four coefficients, $C$, $L_1$, $L_2$, and $D$, extracted from five different experimental data through inverse analysis using Eq. (5). The unit of these coefficients corresponds to pA. The first two digits indicate significant figures. Experimental data are those shown in, from the top row to the bottom row, Fig. 7(a), Fig. 7(b), Fig. 2S, Fig. 3S, Fig. 8(a), and Fig. 4S, respectively. The measurement column represents polarization state at $\varphi_Q = 0°$, power of excitation light, and accumulated illumination time $t$, respectively.

| Sample ID | specification | measurement | $C$ | $L_1$ | $L_2$ | $D$ | $C/D$ | $L_2/D$ |
|---|---|---|---|---|---|---|---|---|
| A | as- cleaved | $p$, 4000 W/cm$^2$, $t$ = 2 hrs | 3.56 | 7.71 | 13.3 | 238 | 0.015 | 0.056 |
| | as- cleaved | $p$, 4000 W/cm$^2$, $t$ = 7 hrs | 1.35 | -3.17 | 6.24 | 145 | 0.009 | 0.043 |
| | polished & $SiO_2$ coated | $p$, 4450 W/cm$^2$, $t$ = 7 hrs | 0.420 | -1.45 | 2.27 | 25.1 | 0.017 | 0.090 |
| B | as-cleaved | $p$, 50 W/cm$^2$, $t$ = 7 hrs | 8.85 | -9.44 | 83.0 | 2660 | 0.003 | 0.031 |
| | polished & $SiO_2$ coated | $p$ 167 W/cm$^2$, $t$ = 5.5 hrs | 0.180 | -0.178 | 0.979 | 12.5 | 0.014 | 0.078 |
| | polished & $SiO_2$ coated | $s$, 167 W/cm$^2$, $t$ = 10 hrs | -0.020 | 0.0126 | -0.225 | 5.55 | -0.004 | -0.041 |

# Supplementary Information for Polarization-dependent photocurrent in a quadrilateral-shaped bulk crystalline tellurium chip with near-infrared light excitation

Hiro Munekata[1]*, Gakuto Kusuno[1], Kohei Miyazaki[1], and Takuya Satoh[1,2]*

[1] *Department of Physics, Institute of Science Tokyo, Tokyo 152-8551, Japan*

[2] *Quantum Research Center for Chirality, Institute for Molecular Science, Okazaki 444-8585, Japan*

* E-mail: munekata.h.aa@m.titech.ac.jp; satoh@phys.sci.isct.ac.jp

## 1. Current-voltage characteristics

Figure 1S shows the current-voltage characteristics of as-cleaved sample A and $SiO_2$ coated sample B measured at room temperature in the dark. Measurements were carried out prior to the photocurrent experiments from which CPGE data shown in Fig. 6(a) and Fig. 7(a) were obtained, respectively. The sample resistance is around 1.6 kΩ along the [001] direction for sample A, whereas, for sample B, they are around 3.1 and 4.6 kΩ along the [001] and [100] directions, respectively. The magnitude relation observed in sample B is qualitatively consistent with that reported earlier in ref. 8.

## 2. Validation of photocurrent measurement setup

Figure 2S shows photocurrent data measured as a function of quarter wave plate (QWP) angle ($\varphi_Q$) with a polycrystalline Cu(In,Ga)$Se_2$ solar cell tilted at the tilt angle $\theta = 45°$. Cells were prepared as described in Ref. 50 by Mr. A. Funaki, a former student member of the Yamada Laboratory in the Electrical and Electronic Engineering Department at the Institute of Science Tokyo. The primary feature of the data is a periodic change in the photocurrent between the LP and CP states, which originates from the polarization-dependent reflection at the cell surface.

Long-term background fluctuation is also noticeable, but not serious. In particular, the fluctuation between the LCP and RCP is small, being $5 \pm 2 \times 10^{-4}$, relative to the photocurrent averaged over one rotation of the QWP. This fact confirms that our measurement setup is nearly free from the artifact caused by circular polarization.

## 3. Photocurrent of regenerated sample A with $SiO_2$ coating

Figure 3S show photocurrent data along the *z* direction of regenerated sample A that was mirror polished and $SiO_2$ coated after deterioration. Recovery of LP-dependent and CP-dependent photocurrents both are clearly observed.

## 4. Photocurrent of as-cleaved sample B

Figure 4S shows photocurrent data obtained from as-cleaved sample B as a function of QWP angle ($\varphi_Q$). Initial polarization at the angle $\varphi_Q = 0°$ is *p* polarization, with light incident angle $\theta = 45°$. Measurement was carried out after accumulated illumination time $t = 7$ h. Excitation wavelength and power density was $\lambda = 780$ nm and 50 W/cm$^2$, respectively. A blue dashed line is a curve calculated with all four terms in Eq. (5). The value of these terms are $C = 0.0089$, $L_1 = -0.0094$, $L_2 = 0.083$, and $D = 2.66$ nA, respectively. The C/D ration is around 0.3 %. The $\beta$ value is estimated to be around 850 nA/W with the window area of a current flow A' = 400 μm × 50 nm.

## 5. Photocurrent of $SiO_2$ coated sample B before deterioration

Figure 5S shows photocurrent data obtained from *$SiO_2$-coated* sample B as a function of QWP angle ($\varphi_Q$). Initial state of polarization is *s* polarization at $\varphi_Q = 0°$. Measurements were carried out after accumulated illumination time of $t = 10$ h, during which the CP dependent characteristics were barely noticeable, suggesting that the sample was just before the state of complete deterioration.

## 6. Inverse analysis with incomplete phenomenological equations

Figure 6S shows the results of the inverse analyses carried out using two differently-modified Eq. (5); the one without a linear PGE, the $L_1$ term in Eq. (5), and the other without a circular PGE, the $C$ term in Eq. (5). Namely, analysis was performed with $I = C\sin 2\varphi_Q + L_2\cos 4\varphi_Q + D$ for the former case, whereas with $I = L_1\sin 4\varphi_Q + L_2\cos 4\varphi_Q + D$, for the latter case. The curve calculated using the former equation shows the disappearance of the slight distortion in the cosine waveform, whereas the curve calculated using the latter equation shows the disappearance of amplitude modulation.

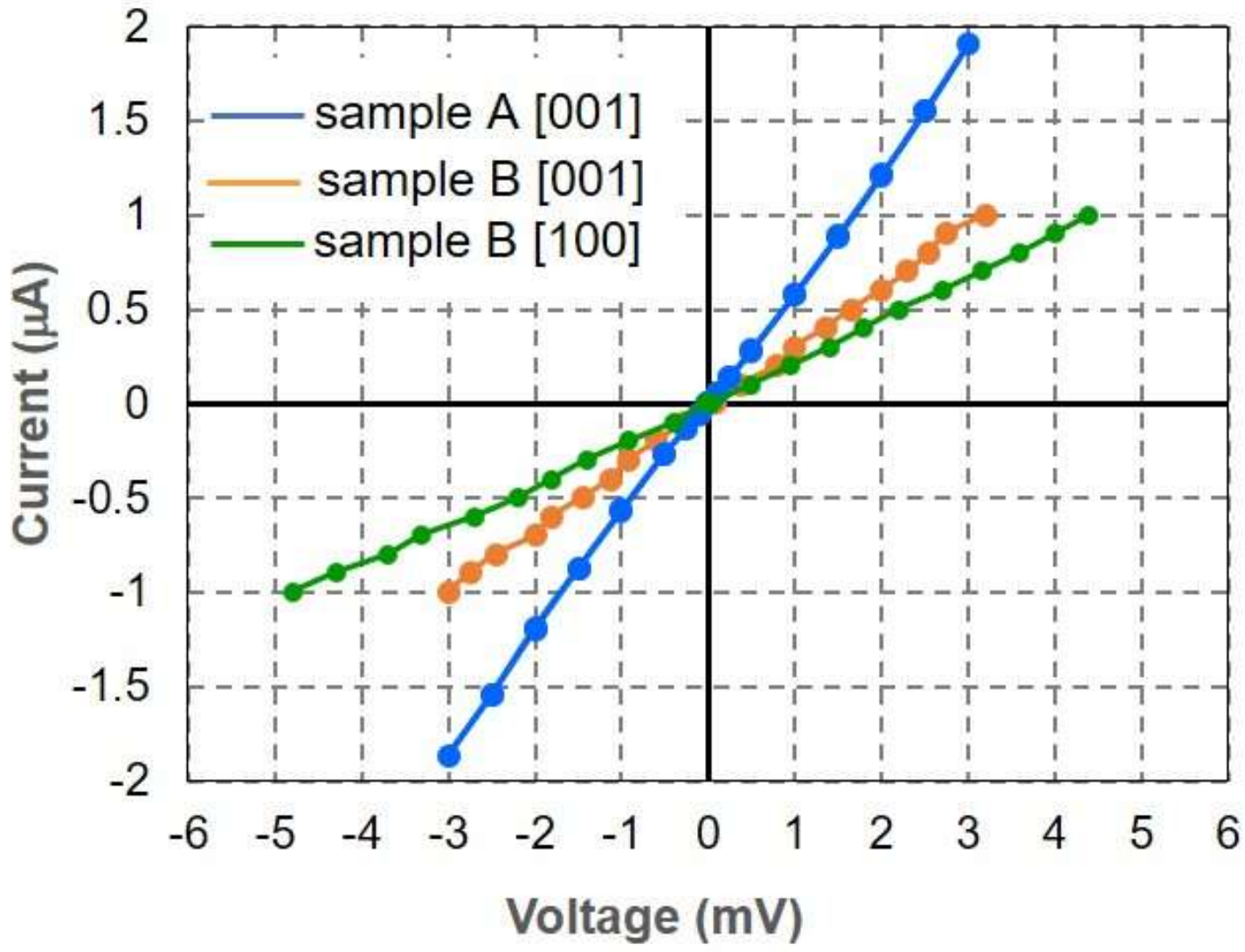

Figure 1S: Current-voltage data of as-cleaved sample A along the [001] direction (a blue line and dots), and of $SiO_2$ coated sample B along the [001] direction (an orange line and dots) and along the [100] direction (a green line and dots). They were obtained at room temperature in the dark.

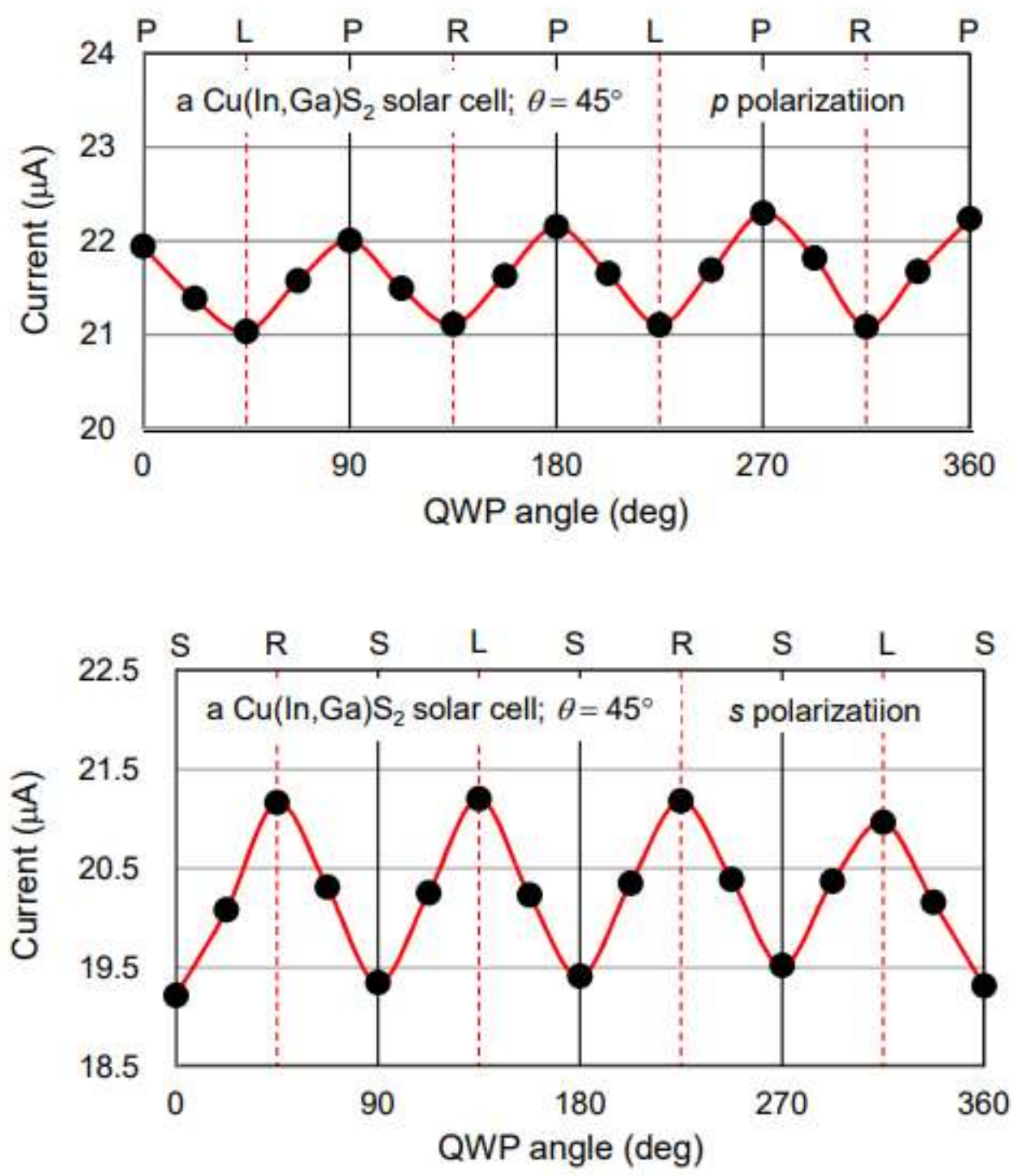

Figure 2S: Plots of photocurrent as a function of quarter-wave plate (QWP) angle ($\varphi_Q$) obtained from a Cu(In,Ga)$Se_2$ solar cell. Initial polarization at the angle $\varphi_Q = 0°$ is (a) $p$ and (b) $s$ polarization, respectively. A laser beam of $\lambda$ = 780 nm and power of 0.037 mW was impinged on the cell mounted with tilt angle of $\theta$ = 45°. Area of illumination was approximately 0.126 $cm^2$. Photocurrent at each $\varphi_Q$ is the average of ten data acquired over 20 sec. The average photocurrent values are 21.7 and 20.2 μA for data shown in (a) and (b), respectively.

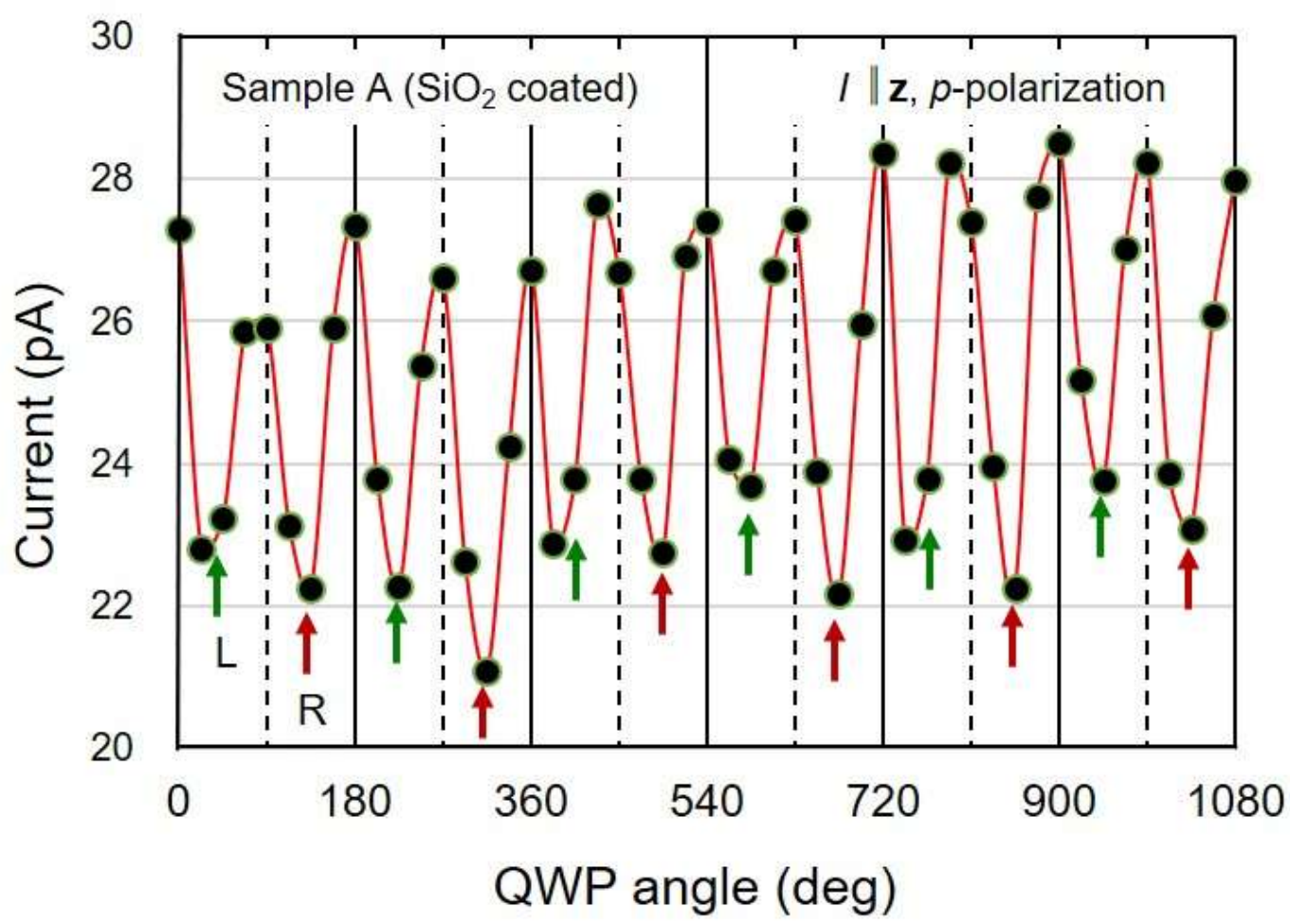

Figure 3S: A plot of photocurrent as a function of quarter-wave plate (QWP) angle ($\varphi_Q$) obtained from regenerated sample A. Measurement was carried out after accumulated illumination time $t$ = 6.5 h. Initial polarization at the angle $\varphi_Q = 0°$ is $p$ polarization, with light incident angle $\theta$ = 45°. Excitation wavelength and power density was $\lambda$ = 780 nm and 4450 W/cm$^2$, respectively. Green and brown arrows with notations L and R represent excitation with LCP and RCP, respectively. A red curved line is for eye guide.

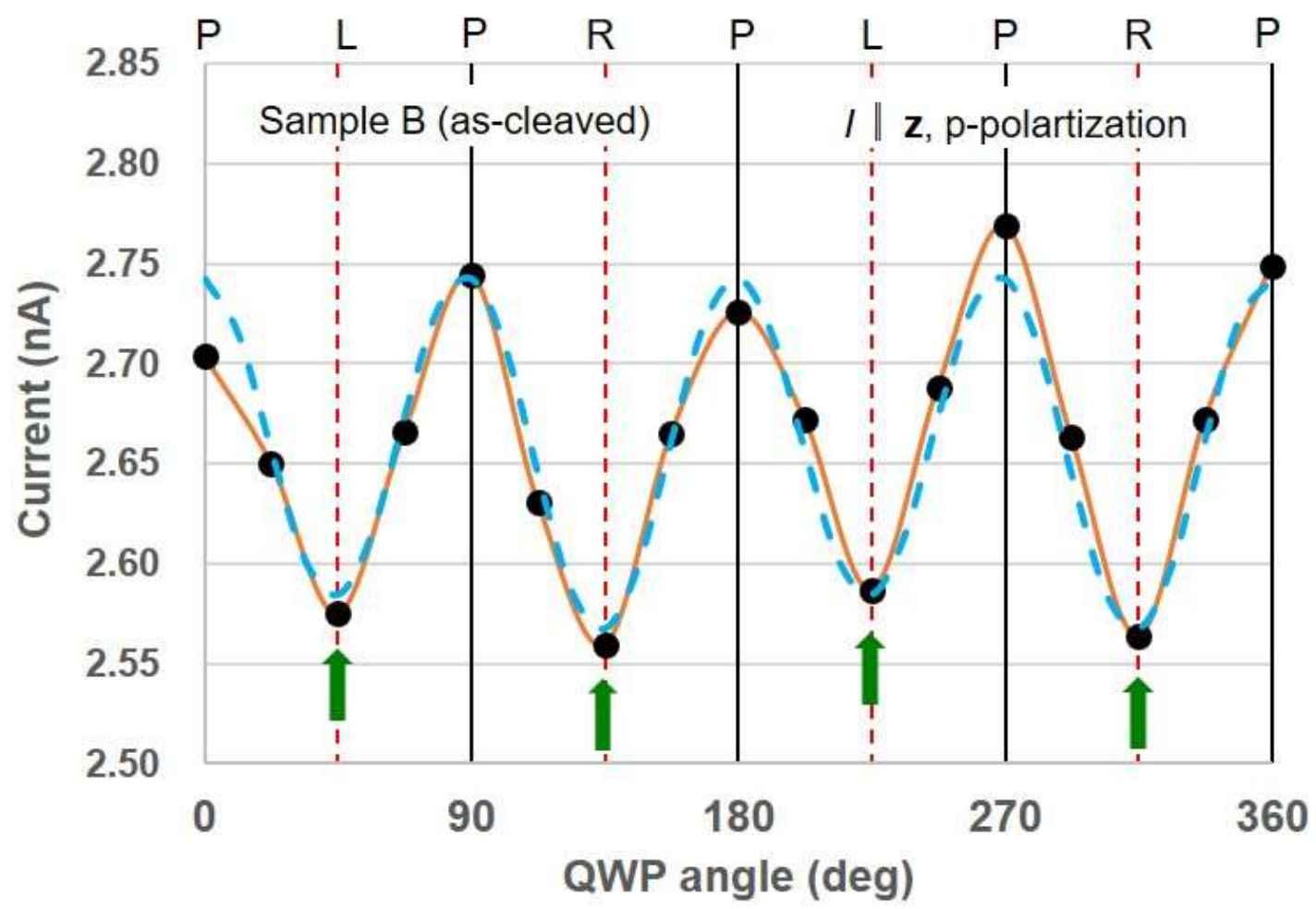

Figure 4S: A plot of photocurrent as a function of quarter-wave plate (QWP) angle ($\varphi_Q$) obtained from as-cleaved sample B. Measurement was carried out after accumulated illumination time $t$ = 7 h. Initial polarization at the angle $\varphi_Q = 0°$ is $p$ polarization, with light incident angle $\theta$= 45°. Excitation wavelength and power density was $\lambda$ = 780 nm and 50 W/cm$^2$, respectively. Green arrows with notations L and R represent excitation with LCP and RCP, respectively. An orange, curved line is for eye guide, whereas a blue, dashed line is a curve calculated with all four terms in Eq. (5).

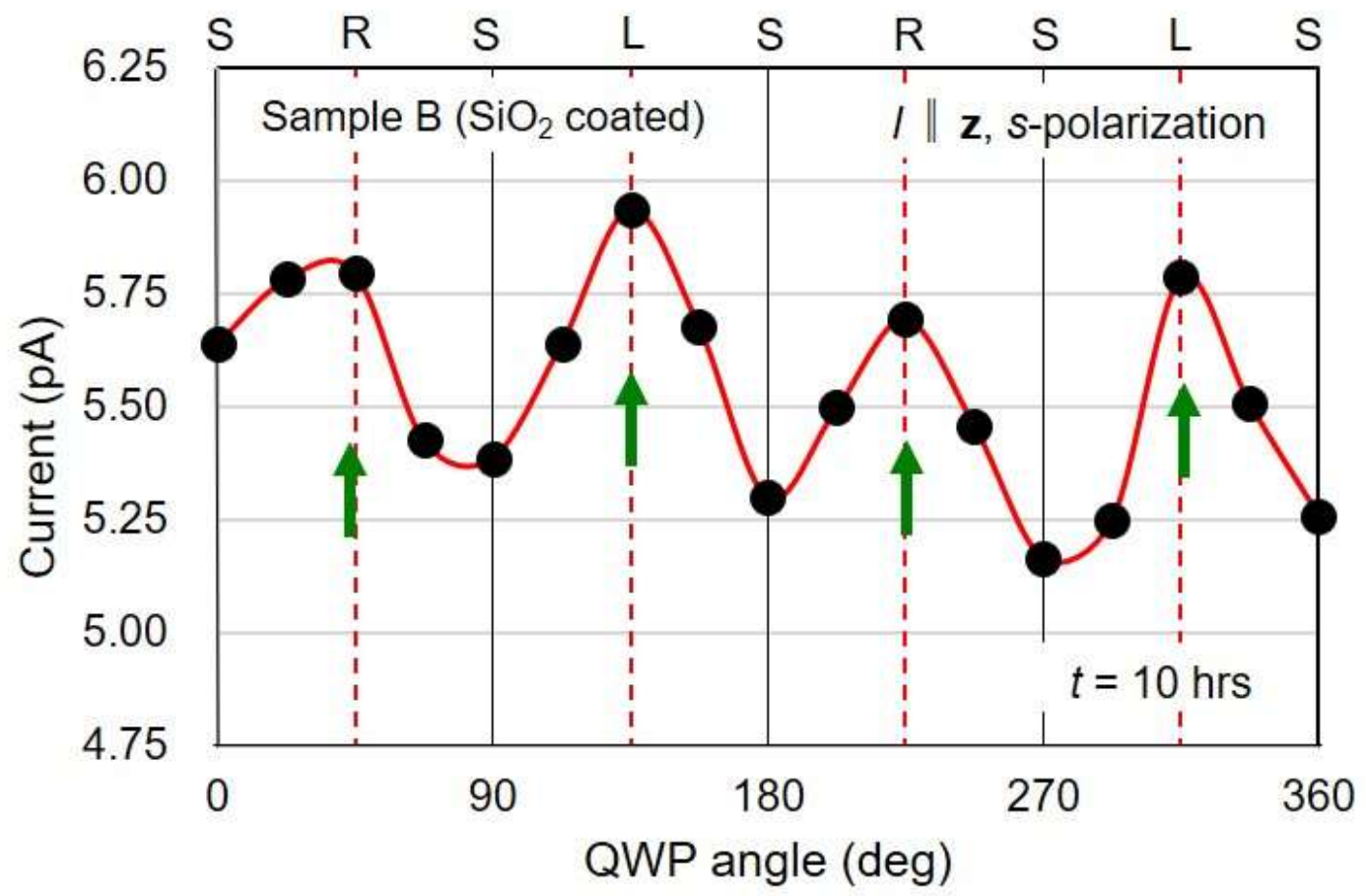

Figure 5S: Photocurrent data obtained from *$SiO_2$-coated* sample B as a function of QWP angle ($\varphi_Q$), with light-incident angles of $\theta$ = 45°, initial polarization of $s$ polarization at $\varphi_Q = 0°$, and excitation power density of 167 W/cm$^2$. Measurements were carried out after total illumination time of $t$ = 10 h. Notations S, R, and L, represent $s$, right-circular, and left-circular polarizations, respectively. Red curved lines are for eye guide.

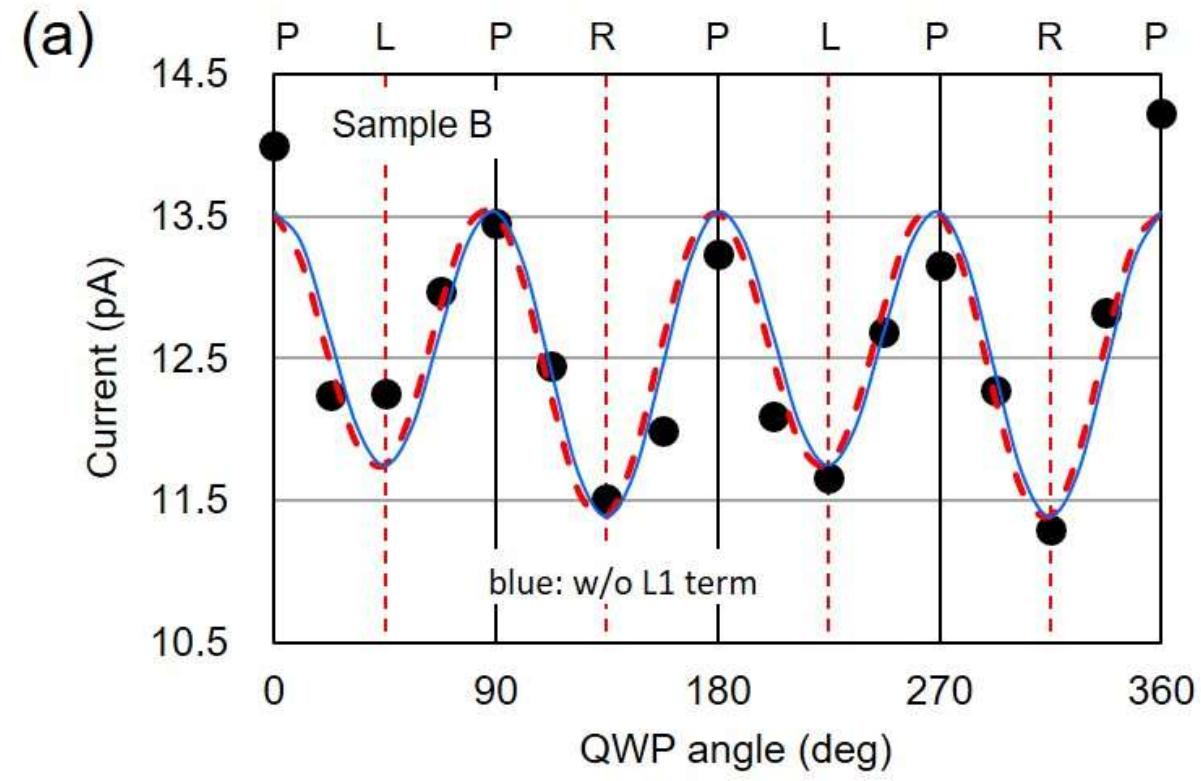

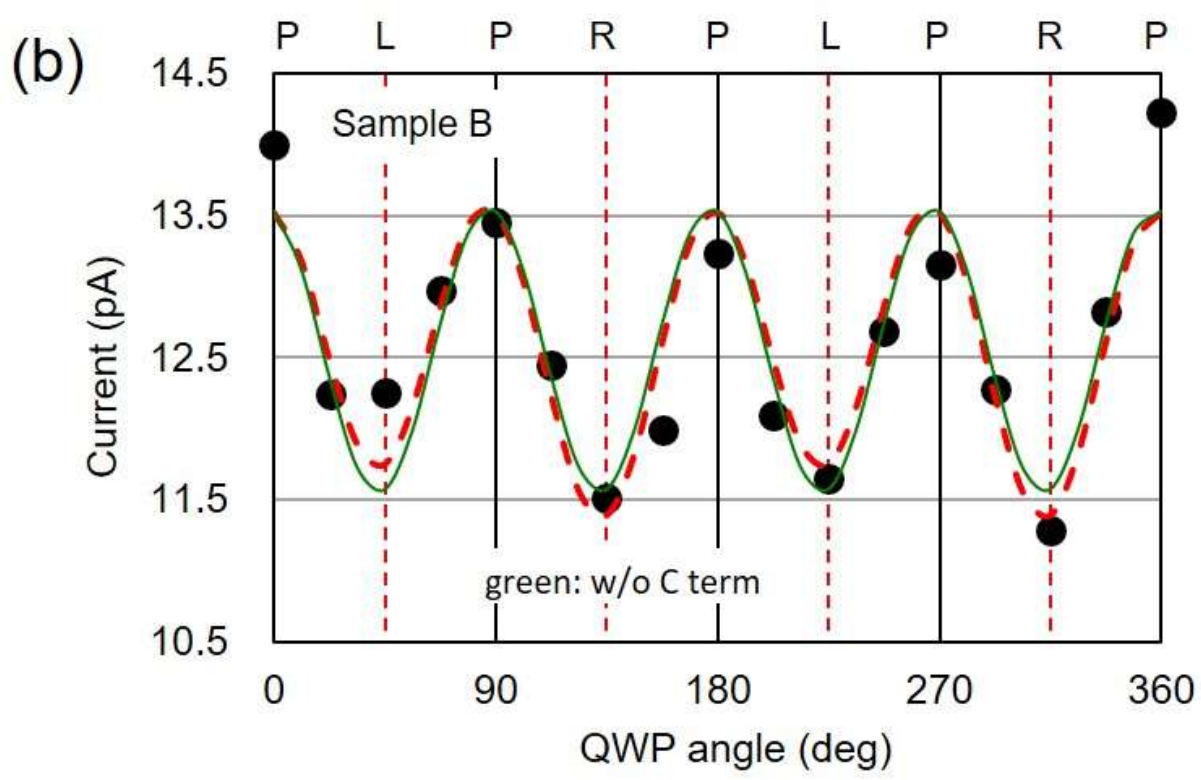

Figure 6S: (a) A curve calculated without the contribution of LPGE (solid blue), the term $L_1$ in Eq. (5), and (b) a curve calculated without the contribution of CPGE (solid green), the C term in Eq. (5). In both panels, black dots represent experimental data shown in Figs. 7(a) and 8(a), whereas red dashed lines are curves calculated with all four terms in Eq. (5).

(END of Supplementary Materials)